\documentclass{article}

\usepackage{arxiv}

\usepackage[utf8]{inputenc} 
\usepackage[T1]{fontenc}    
\usepackage{hyperref}       
\usepackage{url}            
\usepackage{booktabs}       
\usepackage{amsfonts}       
\usepackage{nicefrac}       
\usepackage{microtype}      
\usepackage{lipsum}
\usepackage{graphicx,des,ieee_thm,amsmath,psfrag,url}     
\usepackage{amsmath,array,cite,tikz}
\usepackage{xcolor}
\usetikzlibrary{shapes,arrows}
\def\figscale{.75}
\usepackage{tikz-cd}

\title{Transforming opacity verification to nonblocking verification in modular systems}

\author{
  Sahar Mohajerani \thanks{The work of the first author was supported by the Swedish Research Council. The work of the second author was supported in part by US NSF grant CNS-1421122.} \\
  Department of Electrical engineering and Computer Science\\
  University of Michigan\\
  \texttt{sahar.mohajerani@gmail.com} \\
   \And
   St\'ephane Lafortune\\
   Department of Electrical engineering and Computer Science\\
  University of Michigan\\
  \texttt{stephane@umich.edu} \\
}

\begin{document}
\maketitle

\section{Abstract}
We consider the verification of current-state  and $K$-step opacity for systems modeled as interacting non-deterministic finite-state automata. 
We describe a new methodology for compositional opacity verification that employs abstraction, in the form of a notion called opaque observation equivalence, and that leverages
existing compositional nonblocking verification algorithms.
The compositional approach is based on a transformation of the system, where the transformed system is nonblocking if and only if the original one is current-state opaque. 
Furthermore, we prove that $K$-step opacity can also be inferred if the transformed system is nonblocking. 
We provide experimental results  where current-state opacity is verified efficiently for a large scaled-up system.


\newcommand{\ACTE}{\Sigma_\tau}
\newcommand*{\QEDA}{\hfill\ensuremath{\blacksquare}}%
\section{Introduction} 

While there is a large amount of information people willingly release everyday, there is some information that we wish to remain secret. 
Thus, various notions of security have been studied in the past decades; \emph{opacity} is one such example. 
Opacity is an information flow property that identifies whether or not a secret is released to an external observer of the behavior of a known dynamic system.
We refer to the external observer as the \emph{intruder} in this paper.

The notion of opacity was introduced in the field of discrete event systems in \cite{BrKoRy:05}, where the system is modeled as a Petri net. 
Later, a variety of notions of opacity were introduced to cope with different security requirements. 
\emph{Current-state} opacity \cite{SabHad:07}, \emph{initial-state} opacity \cite{SabHaj:13}, \emph{initial-and-final-state} opacity \cite{WuLaf:13}, \emph{$K$-step} opacity \cite{SabHaji:11tase,FalMar:14},  \emph{infinite-step} opacity \cite{SabHaji:12inf}, and \emph{language-based} opacity \cite{FengLin:11} are some examples of state-based and language-based opacity notions. 
In \cite{WuLaf:13}, polynomial-time algorithms are  presented to transform the verification of current-state, initial-state, initial-and-final-state opacity, and language based-opacity to one another.  
In this paper,  we study the verification of current-state and $K$-step opacity under the framework of modular discrete event systems.
%
A system is said to be current-state opaque if the intruder can never know for sure that the current state of the system is a secret state. 
On the other hand, a system is $K$-step opaque if the intruder cannot determine if the system had entered a secret state within the $K$ previous steps of its observed behavior (i.e., it is a \emph{smoothing} property in system-theoretic terminology).  

In this paper, we consider a class of modular systems that are modeled as partially observed (or non-deterministic) interacting finite state automata. 
The {monolithic} approach to verify any opacity property for modular systems is to synchronize all the components of the system and then use the corresponding verification algorithm on the resulting monolithic system. 
This approach is limited by the well-known {state-space explosion} problem, when composing a large number of system components.

Abstraction and modular approaches are standard techniques that can be used to alleviate the state-space explosion problem, either independently or jointly.
In the opacity verification problem domain, the verification of initial-state opacity in a modular setting was studied in \cite{SabHajWodes:10}, 
where it is shown that the system is initial-state opaque if and only if the strings causing violations of opacity are disabled by synchronization.  
Abstraction-based bisimulation was used in \cite{ZhaZam:17} to reduce the complexity of the system when verifying infinite-step opacity. 
One method to alleviate the state-space explosion problem is the \emph{compositional approach based on abstraction}.  
This approach is well-developed for nonblocking verification and supervisor synthesis in modular systems; see, e.g., \cite{FloMal:09, WarMal:10, MohMalFab:17, MohMalFab:15}. 
The compositional approach seeks to remove and merge states that are redundant for the purpose of verification or synthesis, and it proceeds in an incremental manner in terms of system components.

This paper presents a novel compositional approach for the verification of current-state, infinite-step, and $K$-step opacity. 
As infinite-step opacity is a limiting case of $K$-step opacity, we mainly focus on current-state and $K$-step opacity. 
In our framework, each system component is abstracted using a restricted version of {observation equivalence} or weak bisimulation \cite{Mil:80},  that we call \emph{opaque observation equivalence}. 
After such abstraction, the current-state estimator \cite{SabHaji:08} or the two-way observer \cite{XiangLaf:17} of each component is generated, depending on which opacity property is to be verified, either current-state or $K$-step. 
Next, the opacity verification problem is transformed to a suitable nonblocking verification problem. 
This makes it possible to use well-developed nonblocking verification algorithms to verify the different notions of opacity.  
In the case of current-state opacity, we show that the transformation to nonblocking leads to an equivalent problem, i.e., we show necessity and sufficiency.
In the case of $K$-step  opacity, we show sufficiency of the transformation.
We used the software tool Supremica \cite{AkeFabFloVah:03}  to verify current-state opacity of a large modular system using our compositional approach. 
Specifically, we have successfully verified  current-state opacity for a large system containing 4\E 3 automata under one minute on a standard laptop computer. 

The presentation of our results is organized as follows.
\Sect~\ref{sec:preli} gives a brief background about different notions of opacity. 
\Sect~\ref{sec:compo} explains the general compositional opacity problem. 
Next, Sections~\ref{sec:Compocur} and~\ref{sec:Kstep} explain the compositional approach for current-state and $K$-step opacity, respectively. 
Our experimental results on a scaled-up example are presented in \Sect~\ref{sec:experimental}. 
Finally, some concluding remarks can be found in \Sect~\ref{sec:conclusion}.

\section{Modeling framework}\label{sec:preli} 

\subsection{Automata and their composition}

Discrete system behaviors can be modeled by deterministic or  nondeterministic automata. 

\begin{definition} 
A (nondeterministic) finite-state automaton is a tuple $G= \auttuple{\ACT_{}}$,
where $\Sigma_{}$ is a finite set of events, $Q$ is a finite set of states,
$\intrans \subseteq Q \times \Sigma_{} \times Q$ is the \emph{state
transition relation}, and $\initstateset \subseteq Q$ is the set of
\emph{initial states}.
$G$~is \emph{deterministic} if $|\initstateset| = 1$ and if $x
\trans[\sigma] y_1$ and $x \trans[\sigma] y_2$ always implies that $y_1 = y_2$.
\end{definition}
When marking is important the above definition can be extended to $G=\langle\ACT_{},Q, \trans,Q^\circ, Q^m\rangle$, where $Q^m\subseteq Q$ is the set of \emph{marked states}.  
In this paper, we identify marked states in the figures using gray shading.

We assume that the intruder can only partially observe the system. Thus, $\ACT_{}$ is partitioned into two disjoint subsets, the set of \emph{observable} events and the set of \emph{unobservable} events. Since the identity of unobservable events are irrelevant they are all replaced by a special event $\tau$. The event $\tau$ is never included in the alphabet $\ACT$, unless explicitly mentioned. For this, $\ACT_\tau=\ACT\cup\{\tau\}$ is used \cite{FloMal:09}. Nondeterministic automata hereafter may contain transitions labeled by $\tau$.
However, since $\tau$ represents unobservable events, \emph{deterministic} automata will \emph{never} have $\tau$ transitions. 
In opacity problems, the set of states is also partitioned into two disjoint subsets: $Q^S$ the set of \emph{secret states} and $Q^{NS}=Q \setminus Q^S$ the set of \emph{non-secret states}.

When automata are brought together to interact, lock-step synchronization
in the style of~\cite{Hoa:85} is used.
\begin{definition}
\label{def:synch}
Let $G_1 = \auttuplem[1]{\ACT_{1}}$ and $G_2 = \auttuplem[2]{\ACT_{2}}$ be two nondeterministic automata, with sets of secret states 
$Q_1^S\subseteq Q_1$ and ${Q_2^S}\subseteq Q_2$.
The \emph{synchronous composition} of $G_1$ and $G_2$ is defined as
\begin{equation}
\begin{split}
    G=G_1 \sync  G_2 = \left\langle \Sigma_{1}\cup \Sigma_{2}, Q_1\times Q_2,
                                  \intrans,
                                   \initstateset[1] \times \initstateset[2],
                                   Q_1^m \times Q_2^m
                     \right\rangle 
                                       \end{split}
\end{equation}
where
$$
  \begin{array}{@{}r@{\quad}l@{}}
 		(x_1,x_2) \trans[\sigma] (y_1,y_2) &
    \mbox{if } \sigma \in ({\ACT_1} \cap {\ACT_2}),\\
    & x_1 \trans[\sigma]_1 y_1,\ \text{and}\
    x_2 \trans[\sigma]_2 y_2 \, ; \\
    (x_1,x_2) \trans[\sigma] (y_1,x_2) &
    \mbox{if } \sigma \in (\ACT_1 \setminus \ACT_2)\ 
    \cup\{\tau\}\ \\ & \text{and}\ 
    x_1 \trans[\sigma]_1 y_1 \, ; \\
    (x_1,x_2) \trans[\sigma] (x_1,y_2) &
    \mbox{if } \sigma \in (\ACT_2 \setminus \ACT_1)\ 
    \cup\{\tau\}\ \\&  
    \text{and}\ x_2 \trans[\sigma]_2 y_2 \, ;
  \end{array}
$$
\end{definition}
and where the set of secret states of $G$, $Q^S$, is defined in one of the two following ways:
\begin{enumerate}
\item\label{def:syncland} $Q^S= \{(x_1,x_2)\ | \ x_1\in Q_1^S \ \land \ x_2\in Q_2^S\}$,
\item\label{def:synclor}$Q^S= \{(x_1,x_2)\ | \ x_1\in Q_1^S \ \lor \ x_2\in Q_2^S\}$.
\end{enumerate}
Importantly, this definition of synchronous composition only imposes lock-step synchronization on common events in $\Sigma$.

In the following, whenever necessary, we use the notations $\sync_{\land}$ and $\sync_{\lor}$ to show that the secret states of synchronous composition are defined as in \Defn~\ref{def:synch}~\ref{def:syncland} or~\ref{def:synclor}, respectively. 
When $\sync_\land$ is used, a synchronized state is considered secret if all the composed states are secret. 
In $\sync_{\lor}$ however, if one of the  states of the synchronized state is secret, then the synchronized state is considered secret. $\sync_\lor$ and $\sync_\land$ are the first natural constructs for joint secrecy. 

\ACTstar\ is the set of all finite traces of events from~$\ACT$,
including the \emph{empty trace}~$\varepsilon$.
The
\emph{natural projection} $P_\tau\colon \ACTE^* \to \ACT^*$ is the
operation that removes from traces $t \in \ACT_\tau^*$ all events not in~$\ACT$, which affects only event $\tau$ in our setting. 

 The transition relation of an automaton $G$ is written in infix notation $x \trans[\sigma] y$,
and it is extended to strings in $\Sigma_\tau^*$ by letting $x \trans[\varepsilon]
x$ for all $x\in Q$, and $x\stackrel{t\sigma}{\rightarrow}z$ if
$x\stackrel{t}{\rightarrow}y$ and $y\stackrel{\sigma}{\rightarrow}z$ for
some~$y \in Q$. Furthermore, $x \trans[t]$ means that $x \trans[t] y$ for
some $y \in Q$, and $x \trans y$ means that $x \trans[t] y$ for some $t \in
\ACT_\tau^*$. These notations also apply to state sets, where $X \trans[t]Y$ for $X, Y
\subseteq Q$ means that $x \trans[t]y$ for some $x \in X$ and $y\in Y$, and to automata, where
$G \trans[t]$ means that $\initstateset \trans[t]$, etc.

For brevity, $p\ttrans[s]q$, with $s\in\ACT^*$, denotes the existence of a string $t\in\ACT_{\tau}^*$ such that $P_\tau(t)=s$ and $p\trans[t]q$. Thus, $p\trans[u]q$, $u\in\ACT_\tau^*$, means a path containing exactly the events in $u$, while $p\ttrans[u]q$, $u\in\ACT^*$, means  existence of a path between $p$ and $q$ with arbitrary  number of $\tau$ events between the events of $u$. Similarly, $p\ttrans[\tau] q$ denotes the existence of a string $t\in \{\tau\}^*$ such that $p\trans[t]q$.  

The \emph{language} of an automaton~$G$
is defined as $\LANG(G) = \{\, s \in \ACTstar \mid G \ttrans[s] \,\}$ and the language generated by $G$ from $q\in Q$ is $\LANG(G,q)=\{s\in\ACT^*\ |\ q\ttrans[s]\}$. 
Thus we do not include event $\tau$ in the strings in the language of an automaton.
Moreover, from \Defn~\ref{def:synch}, it follows that $s\in\LANG(G_1\sync G_2)$ if and only if $P_1(s)\in \LANG(G_1)$ and $P_2(s)\in \LANG(G_2)$, 
where $P_i :\ACT_{1}\cup\ACT_{2}\to \ACT_i$, for $i=1,2$  
(and these functions are extended to strings in the usual manner).

Renamings $\Delta\colon \ACT\trans\ACT\times\{\epsilon\}$ and $\Delta_R\colon \ACT\trans\{\epsilon\}\times\ACT$ are two maps such that $\Delta(\sigma)=(\sigma,\epsilon)$ and $\Delta_R(\sigma)=(\epsilon,\sigma)$. Renamings are extended to traces
$s \in \ACTstar$ by applying them to each event, and to languages $L
\subseteq \ACTstar$ by applying them to all traces. They are also
extended to automata with alphabet~$\Sigma$ by replacing all transitions
$x \trans[\sigma] y$ with $x \trans[\Delta(\sigma)] y$ and $x \trans[\Delta_R(\sigma)] y$. 

Given an automaton $G=\langle \ACT_\tau,Q,\trans, Q^\circ\rangle$ the \emph{reversed automaton} of $G$ is a non-deterministic automaton $G_R=\langle \ACTE,Q,\trans_R, Q\rangle$ where $\trans_R=\{(x,\sigma, y)\ | \ y\trans[\sigma] x\}$ and all states are considered to be initial  \cite{WuLaf:13}. 
%
For non-deterministic automaton $G= \auttuple\ACTE$, the set of \emph{unobservably reached states} of $B\in 2^Q$, is $UR(B)=\bigcup\{C\subseteq Q\ |\ B\ttrans[\tau] C \}$. The \emph{observer automaton}  $det(G)=\langle \ACT, X_{obs},\trans_{obs},X_{obs}^\circ\rangle$ is a deterministic automaton, where $X_{obs}^\circ=UR(Q^o)$ and $X_{obs}\subseteq 2^Q$, and $X\trans_{obs}Y$, where $X, Y\subseteq X_{obs} $, if and only if $Y=\bigcup\{UR(y)\ | \ x\trans[\sigma] y\ \textnormal{for some}\ x\in X \ \textnormal{and}\ y\in Q \}$. By convention, in this paper only reachable states from $X^\circ_{obs}$ under $\trans_{obs}$ are considered in $X_{obs}$. Also, we will refer to the observer automaton as the \emph{current-state estimator}, abbreviated as \emph{CSE}.

In this paper, the special blocking event $\psi\notin \ACT$ is used to label additional transitions going out of a special set of states, termed $\psi$-states and denoted by $X_\psi$. 
\begin{definition}\label{def:omegaAutomaton}
Let $G=\langle\ACT,Q, \trans,q^\circ\rangle$  be a deterministic automaton with set of $\psi$-states $X_\psi\subseteq Q$. Then $G_\psi=\langle\ACT\cup\{\psi\},Q\cup\{\dumpstate\}, \trans_\psi,q^\circ,Q^m\rangle$ is a deterministic automaton  such that $\dumpstate$ is a new state,   $Q^m=Q$, which means that all the original states are marked, and 
\begin{equation}
\trans_\psi = \trans \cup \{ (x,\psi,\dumpstate)\ | \ x\in X_\psi \}
\end{equation} 
\end{definition}
This ``$\psi$-transformation'' will be used later on to transform opacity verification to nonblocking verification. To check if a specific state of the system can be reached, the state can be considered as a $\psi$-state and nonblocking verification can be done on the transformed system, termed the $\psi$-system. In our setting   the  $\psi$-states are the states that violate opacity.  

Another common automaton operation is the \emph{quotient} modulo an
equivalence relation on the state set.

\begin{definition}
Let $Z$ be a set. A  relation $\insim \subseteq  Z\times Z$ is called an
\emph{equivalence relation} on~$Z$ if it is reflexive,
symmetric, and transitive.
Given an equivalence relation~\insim\ on~$Z$, the \emph{equivalence class} of $z \in Z$
is $[z] = \{\, z' \in Z \mid z \sim z' \,\}$,
and $\tilde{Z} = \{\, [z]
\mid z \in Z \,\}$ is the set of all equivalence classes modulo~$\sim$.
\end{definition}

\begin{definition}
Let $G = \auttuple\ACTE$ be an automaton and let $\insim \subseteq Q \times
Q$ be an equivalence relation. The \emph{quotient automaton} of $G$
modulo~$\sim$ is
\begin{equation}
  \tilde{G} = \langle \Sigma_{\tau}, \tilde{Q}, \intrans\modsim,
                     \tilde{Q}\init \rangle \ ,
\end{equation}
where $\intrans\modsim = \intrans\modsim =\{\, ([x],\sigma,[y]) \mid \exists x'\in [x], y'\in[y]:x' \trans[\sigma] y'
\,\}$ and $\tilde{Q}\init =\{\, [x\init] \mid x\init \in
\initstateset \,\}$.
\end{definition}

\subsection{Notions of opacity}

In general, opacity addresses the issue whether an intruder observing the system, and knowing the model of the system, can determine for sure if the system is in a secret state. There are different notions of opacity in the literature. It is shown in~\cite{WuLaf:13} that initial, final, current-state, and language-based opacity can be transformed to one another with polynomial-time  algorithms for verification purposes. Moreover, infinite-step opacity is a limiting case of $K$-step opacity. Thus, in this paper, we mainly address the verification of current-state opacity first, and then that of $K$-step opacity, which cannot be transformed to current-state opacity for verification purposes.
We recall the formal definitions of these properties in the context of the framework of this paper.
\begin{definition}\label{def:currentStateOp}
A non-deterministic automaton $G$ with event set $\ACT$ and set of secret states $Q^S$ is \emph{current-state opaque}, with respect to $Q^S$ if and only if
$$
\begin{array}{@{}r@{\quad}l@{}}
&(\forall q_0\in Q^o, \forall s\in\LANG(G,q^\circ) \ \colon\ q^\circ\ttrans[s]Q^S)\\
& (\exists q'^\circ\in Q^o)\ \textit{such~that}\ [\ q'^\circ\ttrans[s]Q^{NS}]
\end{array}
$$
\end{definition}
The system is current-state opaque if an intruder cannot determine whether the system is currently in a secret state or not.

\begin{definition}\label{def:infinitStateOp}
A non-deterministic automaton $G$ with event set $\ACT$ and set of secret states $Q^S$ is \emph{infinite-step opaque}, with respect to $Q^S$ 
$$
\begin{array}{@{}r@{\quad}l@{}}
&(\forall q_0\in Q^o, \forall st\in\LANG(G,q^\circ) \ \colon\ q^\circ\ttrans[s]Q^S)\\
& (\exists q'^\circ\in Q^o)\ \textit{such~that}\ [\ q'^\circ\ttrans[s]Q^{NS}\land q'^\circ\ttrans[st]]
\end{array}
$$
\end{definition}
The system is infinite-step opaque if an intruder cannot determine whether the system ever was in a secret state or not at any time in the past. 

\begin{definition}\label{def:kstepOp}
A non-deterministic automaton $G$ with event set $\ACT$ and  set of secret states $Q^S$ is \emph{$K$-step opaque}, with respect to $Q^S$ 
$$
\begin{array}{@{}r@{\quad}l@{}}
&(\forall q_0\in Q^o, \forall st\in\LANG(G,q^\circ) \ \colon\ q^\circ\ttrans[s]Q^S\land |t|\leq K)\\
& (\exists q'^\circ\in Q^o)\ \textit{such~that}\ [\ q'^\circ\ttrans[s]Q^{NS}\land q'^\circ\ttrans[st]]
\end{array}
$$
\end{definition}
The system is $K$-step opaque if the entrance of the system into a secret state remains uncertain for an intruder after up to $K$ future observations.
Hence, 0-step opacity is equivalent to current-state opacity,
and when $K\trans\infty$, then $K$-step opacity becomes infinite-step opacity~\cite{SabHaji:12inf}. 

It is shown in~\cite{SabHaji:08} that current-state opacity can be verified by building the standard observer automaton. 
%
\begin{proposition}\label{def:currentStateOpEstimator}\cite{SabHaji:08}
Let $G=\auttuple{\ACTE}$ be a non-deterministic automaton with  set of secret states $Q^S$.  Let  $det(G) = \langle \ACT,X_{obs},\trans_{obs},X_{obs}^\circ\rangle$ be the current-state estimator of $G$. Then $G$  is current-state opaque with respect to $Q^S$ if and only if for all $s\in\LANG(G)$ it holds that $[det(G)\trans[s]X \ \text{implies that}\  X\not\subseteq Q^S]$.
\end{proposition}

The verification of infinite-step and $K$-step opacity is considered in \cite{SabHaji:11tase} and \cite{SabHaji:12inf}, respectively, where these properties were first introduced.
Recently, a new approach for the verification of infinite and $K$-step opacity was introduced, which relies on building the so-called \emph{two-way observer} of the system  \cite{XiangLaf:17}.
We will leverage this latter approach in the development of our results.
Again, we recall relevant definitions and results, restated in the context of our framework.
\begin{definition}\label{def:twObs}\cite{XiangLaf:17}
Let $G=\auttuple{\ACTE}$ be a non-deterministic automaton and $G_R$ be the reversed automaton of $G$. Let $\Delta\colon \ACT\trans\ACT\times\{\epsilon\}$ and $\Delta_R\colon \ACT\trans\{\epsilon\}\times\ACT$. The \emph{two-way} observer of $G$ is the deterministic automaton obtained by: 
$$H=\Delta(det(G))\sync \Delta_R(det(G_R))$$
\end{definition}

\begin{proposition}\label{def:infinitStateOpTWObs}\cite{XiangLaf:17}
Let $G=\auttuple{\ACTE}$ be a non-deterministic automaton with set of secret states $Q^S$. Let $\Delta\colon \ACT\trans\ACT\times\{\epsilon\}$ and $\Delta_R\colon \ACT\trans\{\epsilon\}\times\ACT$ and let $H=\langle \Delta(\ACT)\cup\Delta_R(\ACT),Q_H,\trans_H,Q^\circ_H\rangle$ be the two-way observer of $G$. Then $G$ is \emph{infinite-step opaque} with respect to $Q^S$  if and only if  
\begin{equation}
H\trans[s](h_1,h_2) \ \text{implies}\  [ h_1\cap h_2\not\subseteq Q^S \lor h_1\cap h_2=\emptyset]
\end{equation}
\end{proposition}
The system is infinite-step opaque if for all the reachable states of the two-way observer, the intersection of the first with the second components is not a subset of the secret states of the system or it is empty. 

\begin{proposition}\label{def:kStateOpTWObs}\cite{XiangLaf:17}
Let $G=\auttuple{\ACTE}$ be a non-deterministic automaton with set of secret states $Q^S$. Let $\Delta\colon \ACT\trans\ACT\times\{\epsilon\}$ and $\Delta_R\colon \ACT\trans\{\epsilon\}\times\ACT$ and let $H=\langle \ACT_H,Q_H,\trans_H,Q^\circ_H\rangle$, with $\ACT_H=\Delta(\ACT)\cup\Delta_R(\ACT)$, be the two-way observer of $G$. Let $P_\Delta:\ACT_H\trans \Delta_R(\ACT)$. Then $G$ is $K$-step opaque with respect to $Q^S$ if and only if, for any string $s\in\LANG(H)$ such that $Q^\circ_H\trans[s](h_1,h_2)$, we have that 
$$
\begin{array}{@{}r@{\quad}l@{}}
[\ h_1\cap\ h_2\subseteq Q^S \land\ h_1\cap h_2\not =\emptyset\ ]\  \Rightarrow\ |P_\Delta(s)|> K.
\end{array}
$$
\end{proposition}

\begin{figure}
	\centering
		\includegraphics{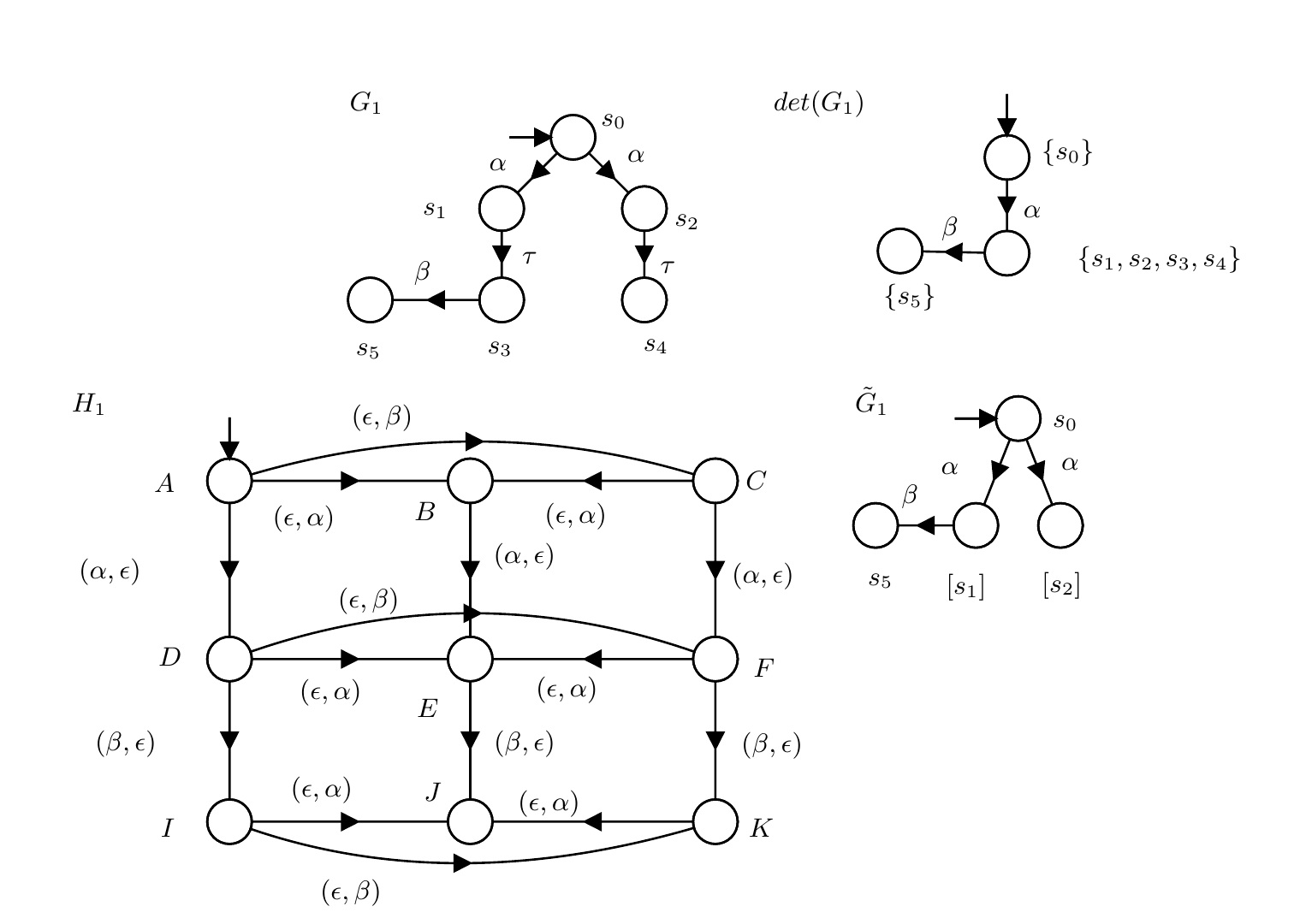}
	\caption{Automata of \Examp~\ref{ex:intro}}
	\label{fig:notion}
\end{figure}

\begin{example}\label{ex:intro}
Consider automaton $G_1$ in \Fig~\ref{fig:notion}. Assume that the set of secret states is $Q^S=\{s_1,s_3\}$; we have that $\Delta\colon\ \{\alpha,\beta\}\to\ \{(\alpha,\epsilon),(\beta,\epsilon)\}$ and $\Delta_R\colon\ \{\alpha,\beta\}\to\ \{(\epsilon,\alpha),(\epsilon,\beta)\}$. In the figure $det(G_1)$ and $H_1$ are the current-state estimator and the two-way observer of $G_1$,  respectively. In $H_1$ the states are:
\begin{itemize}
	\item $A=(\{s_0\},\{s_0,s_1,s_2,s_3,s_4,s_5\})$,
	\item $B=(\{s_0\},\{s_0\})$,
	\item $C=(\{s_0\},\{s_1,s_3\})$,		
	\item $D=(\{s_1,s_2,s_3,s_4\},\{s_0,s_1,s_2,,s_3,s_4,s_5\})$,
	\item $E=(\{s_1,s_2,s_3,s_4\},\{s_0\})$,		
	\item $F=(\{s_1,s_2,s_3,s_4\},\{s_1,s_3\})$,		
	\item $I=(\{s_5\},\{s_0,s_1,s_2,s_3,s_4,s_5\})$,		
	\item $J=(\{s_5\},\{s_0\})$,		
	\item $K=(\{s_5\},\{s_1,s_3\})$.
\end{itemize}
The system is current-state opaque as the intruder cannot distinguish between $s_0\trans[\alpha]s_1$ and $s_0\trans[\alpha]s_2$. Automaton $det(G_1)$ confirms that the system is current-state opaque since there is no state in $det(G_1)$ which is subset of $Q^S$ (\Propn~\ref{def:currentStateOpEstimator}). However, the system is not infinite-step opaque,  because after observing event $\beta$ the intruder will know that system was in $s_1$ and $s_3$. The two-way observer $H_1$ confirms this result as in the state $F$ we have $\{s_1,s_2,s_3,s_4\}\cap\{s_1,s_3\} = \{s_1,s_3\}\subseteq Q^S$ (\Propn~\ref{def:infinitStateOpTWObs}). The system is not $K=1$ step opaque either since $H_1\trans[(\epsilon,\beta)(\alpha,\epsilon)]F$ and $\{s_1,s_2,s_3,s_4\}\cap\{s_1,s_3\} = \{s_1,s_3\}\subseteq Q^S$ but  $P_\Delta((\epsilon,\beta)(\alpha,\epsilon))=(\epsilon,\beta)$ and $|P_\Delta((\epsilon,\beta)(\alpha,\epsilon))|\not> 1$ (\Propn~\ref{def:kStateOpTWObs}).
\end{example}

\section{Compositional opacity verification}\label{sec:compo} 
This section describes the general framework of transforming current-state opacity and $K$-step opacity  to nonblocking verification. 
Since infinite-step opacity is a limiting case of $K$-step opacity, a specific treatment of this property is omitted hereafter; instead, we make relevant observations about it in our discussion; see \Sect~\ref{sec:Kstep}. 
Note that current-state opacity is also a special case of $K$-step opacity. 
However, since verification of current-state opacity requires building the current state estimator and not the two-way observer, we address current-state opacity separately from $K$-step opacity.

The input to the algorithm is a modular nondeterministic system. 
A modular system is a collection of interacting components
\begin{equation}
  \label{plant}
  \SYSG  = G_1 \sync \cdots \sync G_n \ .
\end{equation}
The compositional opacity verification algorithm is summarized in \Fig~\ref{fig:composAlg} and the steps are as follow:
\begin{enumerate}
\item At the first of the compositional opacity verification, the modular
system~(\ref{plant}) is abstracted, using \emph{opaque observation equivalence}.
Each automaton $G_i$ may be replaced by an abstracted version, 
$\tilde{G}_i$, with less states or transitions.

\item Next, the current-state estimators, in the case of current-state opacity verification, or the two-way observers, in the case of   $K$-step opacity verification, of the individual abstracted components are built, $H_i$ in \Fig~\ref{fig:composAlg}.

\item Next, the opacity verification problem is transformed to nonblocking verification problem. The states of the individual current-state estimators or the two-way observers that violate opacity are identified and transitions to blocking states from those states are added, resulting in $H_{i,\psi_i}$  in \Fig~\ref{fig:composAlg}. 

\item Compositional nonblocking verification is used to verified opacity problem. In compositional  nonblocking verification, the synchronous composition is computed gradually, abstracting each intermediate result again.  
Eventually, the procedure leads to a single automaton, denoted by~$\tilde{H}$,  which due to the abstraction process
has less states and transitions compared to the original system. 
Once $\tilde{H}$ is found, it is used for nonblocking verification. The system is current-state opaque if  and only if $\tilde{H}$ is nonblocking  and it is  $K$-step opaque if $\tilde{H}$ is nonblocking.
\end{enumerate}

Our motivation for proceeding as above is that compositional nonblocking verification has been well studied and it has shown very promising results \cite{FloMal:09,WarMal:10}.   

The monolithic approach to verify opacity, first synchronizes all the component of the system and builds the monolithic  current-state estimator or two-way observer of the system. 
%
%
As the number of the states of the synchronized product grows exponentially with the number of components, the complexity of building  the CSE or the two-way observer of the whole system is  $\mathcal{O}(2^{|X|^n})$. In contrast, the complexity of generating modular CSEs or two-way observers, instead of their monolithic counterparts, is $\mathcal{O}(2^{|X|})$, which is significantly smaller. 
In addition, the proposed approach in this paper  not only avoid building synchronized product of the whole system, but it also abstracts the components and reduces the number of the states of each component before the construction of CSEs or two-way observers.  
\Fig~\ref{fig:composAlg} illustrates the steps of compositional opacity verification for the two cases of $\sync_\lor$ and $\sync_\land$.
The subsequent sections formally develop this approach.

\tikzstyle{block} = [rectangle, draw, fill=gray!20, 
    text width=10em, text centered, rounded corners, minimum height=1.9em]
\tikzstyle{line} = [draw, -latex']
\tikzstyle{cloud} = [draw, rounded rectangle ,fill=gray!20,
    text width=22em,]
    \begin{figure}
		\centering
\begin{tikzpicture}[node distance = 2.7cm, auto]
    \node [block] (init) {$input: G_1\sync\ldots\sync G_n$};
    \node [block, below of=init] (ooe) {$\tilde{G}_1\sync\ldots\sync \tilde{G}_n$};
    \node [block, below of=ooe] (det) {$H_1\sync\ldots\sync H_n$};
    \node [block, below of= det] (psi) {$ H_{1,\psi_1} \sync  \ldots\sync H_{n,\psi_n}$};
    \node [cloud, below of=psi] (final) {The system is current-state opaque if and only the result is nonblocking and $K$-step opaque if the result is nonblocking};
    \path [line] (init) -- node [anchor=center, text width=3cm, left, midway]{opaque observation equivalence} node [anchor=center, text width=4.5cm, right, midway] { \Thm~\ref{thm:OpaqueOECS}, Corollary~\ref{cor:obsEqCSLand}, \Thm~\ref{them:obseqInfinit}, Corollary~\ref{cor:obseqInfinit}, \Thm~\ref{them:KstepOE}, Corollary~\ref{cor:KstepOE}} (ooe);
    \path [line] (ooe) -- node [anchor=center, text width=3.5cm, left, midway]{current-state estimator or two-way observer} node [anchor=center, text width=4cm, right, midway]{\Thm~\ref{thm:syncLorCO}, Corollary~\ref{cor:syncLandCO}, \Thm~\ref{theom:synctwoInfinit}, Corollary~\ref{col:synctwoInfinit} \Thm~\ref{them:syncTwoK}, Corollary~\ref{col:syncTwoKland}}(det);
    \path [line] (det) -- node [anchor=center, text width=3.5cm, left, midway] {transforming opacity to} node [anchor=center, text width=4cm, right, midway] { nonblocking verification}(psi);
    \path [line] (psi) -- node[anchor=center, text width=3.5cm, left, midway]  {nonblocking verification} node [anchor=center, text width=4cm, right, midway]{\Thm~\ref{them:OpaqueToBlockigCSLor}, Corollary~\ref{them:OpaqueToBlockigCSLand}, \Thm~\ref{them:infiniteBlockingLor}, Corollary~\ref{cor:infiniteBlockingLand},\Thm~\ref{them:OpBlockingKlor}, Corollary~\ref{cor:OpBlockingKland}}(final);
\end{tikzpicture}
		\caption{Compositional opacity verification procedure.}\label{fig:composAlg}
\end{figure}
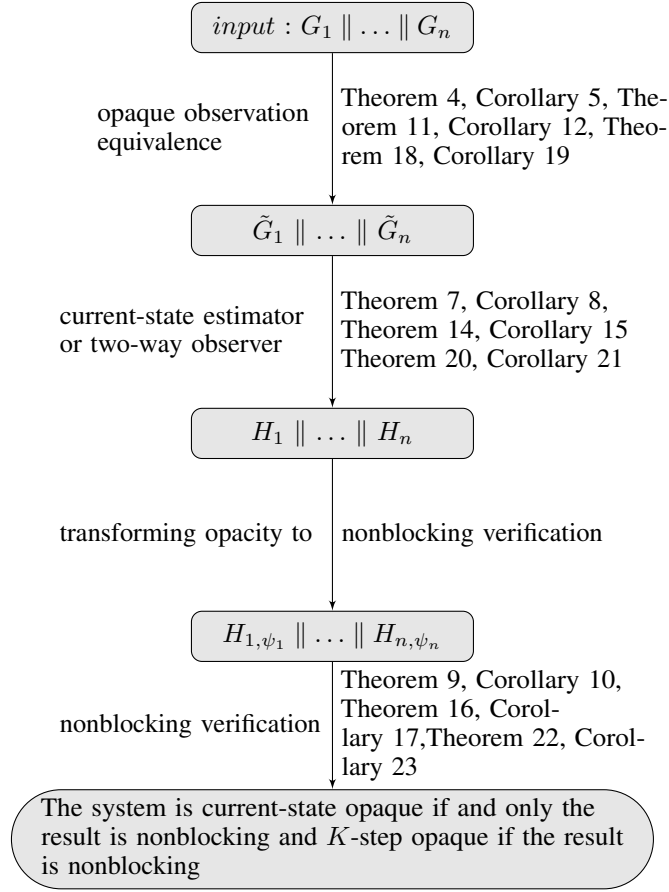

\section{Compositional current-state opacity verification}\label{sec:Compocur} 

This section describes compositional current-state opacity verification. First, \Sect~\ref{sec:obsEqCurrent} describes the abstraction methods that preserve current-state opacity. Next, \Sect~\ref{sec:compos} describes that individual current-state estimators can be built instead of the monolithic current-state estimator. Finally, \Sect~\ref{sec:currentNon} explains the transformation of current-state opacity verification to compositional nonblocking verification. 

\subsection{Opaque observation equivalence}\label{sec:obsEqCurrent}

At the first stage of  compositional opacity verification, individual nondeterministic components are replaced by their abstracted  opaque equivalent components, step (i) in \Fig~\ref{fig:composAlg}.

\emph{Bisimulation equivalence} and \emph{observation equivalence} are two well-known abstraction methods~\cite{Mil:89} to abstract the state space of an automaton. Bisimulation considers states to be equivalent if they have  outgoing transitions with the same events, including unobservable events, to equivalent states.  Observation equivalence is more general than bisimulation as it ignores the unobservable events (namely, event $\tau$ in our set-up).

\begin{definition}\label{def:obseqgeneral}\cite{Mil:89}
  Let $G=\auttuple{\ACTE}$ be a non-deterministic automaton.
  An equivalence relation $\approx\ \subseteq Q \times Q$ is called an
  \emph{ observation equivalence} on~$G$,  if the
  following holds for all $x_1,x_2 \in Q$ such that $x_1 \approx x_2$:
 if  $x_1 \ttrans[s] y_1$ for some $s \in \ACTstar$, then there
  exists $y_2 \in Q$  such that $x_2 \ttrans[s] y_2$, and $y_1\approx y_2$.
\end{definition}

In order to use observation equivalence for abstraction in our compositional opacity verification methodology, the set of secret states needs to be taken into account. For this purpose, a restricted version of observation equivalence called \emph{opaque observation equivalence} is defined.

\begin{definition}\label{def:obseq}
  Let $G=\auttuple{\ACTE}$ be a non-deterministic automaton with  set of secret states $Q^S\subseteq Q$ and  set of non-secret states $Q^{NS}=Q\setminus Q^S$.
  An equivalence relation $\inobseq \subseteq Q \times Q$ is called an
  \emph{opaque observation equivalence} on~$G$ 
  if the following holds for all $x_1,x_2 \in Q$ such that $x_1\obseq x_2$:
\begin{enumerate}
  \item if
  $x_1 \ttrans[s] y_1$ for some $s \in \ACTstar$, then there
  exists $y_2 \in Q$  such that $x_2 \ttrans[s] y_2$, and $y_1\obseq y_2$,
  \item $x_1\in Q^S$ if and only if $x_2\in Q^S$.
\end{enumerate}  
\end{definition}
Opaque observation equivalence considers two states to be equivalent if they have the same secret property and from both of them equivalent states can be reached by the same sequences of events aside from the $\tau$ event. 

We present our first result on the use of opaque observation equivalence in the verification of opacity.
(In the sequel, for the sake of simplicity of notation, we will denote the event set of non-deterministic automata by $\ACT$, with the understanding that some transitions may be labeled by $\tau$.)

\def\OpaqueOECS{%
  Let $\mathcal{G}=\{G_1,\ldots, G_n\}$ be a non-deterministic system with $\sync_\lor$ for interaction, where each automaton has  set of secret states $Q_i^S$. Hence, the set of secret states of the system is $Q^S=Q\setminus Q^{NS}$, where $Q^{NS}=Q^{NS}_1\times\ldots\times Q^{NS}_n$.  Let $\sim$ be an opaque observation equivalence on $G_1$ such that $\tilde{\SYSG}=\{\tilde{G_1},G_2,\ldots,G_n\}$. Then $\SYSG$ is current-state opaque if and only if $\tilde{\SYSG}$ is current-state opaque. }

\begin{theorem}\label{thm:OpaqueOECS}
\OpaqueOECS
\end{theorem}

\emph{Proof:} 

Consider  $T=G_2\sync G_3\sync \ldots\sync G_n=\langle \ACT_T,Q_T,\trans,Q^\circ_T\rangle$. Then $Q^S_T=Q^T\setminus Q_T^{NS}$, where $Q_T^{NS}=Q^{NS}_2\times\ldots\times Q^{NS}_n$. Let $P_G:\ACT_T\cup \ACT_{G_1}\to\ACT_{G_1}$ and $P_T:\ACT_T\cup \ACT_{G_1}\to\ACT_T$. It suffices to show that if $G_1\sync_\lor T$ is not current-state opaque then $\tilde{G_1}\sync_\lor T$ is not current-state opaque either, and vice versa. 
\begin{enumerate}
\item Assume that $G_1\sync_\lor T$ is not current-state opaque. Then there exists $(x_1,x_T)\in Q_{G_1}\times Q_T$ such that $G_1\sync_\lor T\ttrans[s](x_1,x_T)$ and $x_1\in Q_1^S$ or $x_T\in  Q_T^S$, and there does not exist $(x'_1,x'_T)\in Q^{NS}_1\times Q^{NS}_T$ such that $G_1\sync_\lor T\ttrans[s](x'_1,x'_T)$. From $G_1\sync_\lor T\ttrans[s](x_1,x_T)$ it follows that $G_1\ttrans[P_G(s)]x_1$ and $ T\ttrans[P_T(s)]x_T$. 

From  there does not exist $(x'_1,x'_T)\in Q^{NS}_{1}\times Q^{NS}_T$ such that ${G_1}\sync_\lor T\ttrans[s](x'_1,x'_T)$, it follows that for all $x_1'\in Q_{G_1}$ such that $G_1\ttrans[P_G(s)]x'_1$ it holds that  $x'_1\in Q_1^{S}$ or for all $x_T'\in Q_{T}$ such that $T\ttrans[P_T(s)]x'_T$ it holds that  $x'_T\in Q_T^{S}$. 
Moreover, since $G_1$ and $\tilde{G_1}$ are opaque observation equivalent from  $G_1\ttrans[P_G(s)]x_1$ and based on  \Defn~\ref{def:obseq}, it follows that $\tilde{G_1}\ttrans[P_G(s)][x_2]$ such that $x_1\in [x_2]$. 
Now, consider three cases:
\begin{enumerate}
\item\label{item:obseq1} $x_T\in Q_T^S$ and $x_1\in Q_1^{NS}$. Then from $\tilde{G_1}\ttrans[P_G(s)][x_2]$ and $ T\ttrans[P_T(s)]x_T$, it follows that $\tilde{G_1}\sync T\ttrans[s]([x_2],x_T)$. State $([x_2],x_T)$ is also considered secret as $x_T\in Q_T^S$. Since for all $x_T'\in Q_{T}$ such that $T\ttrans[P_T(s)]x'_T$ it holds that  $x'_T\in Q_T^{S}$, then $\tilde{G_1}\sync_\lor T\ttrans[s]([x_2],x'_T)$ such that $x'_T\in Q_T^{NS}$ does not exist. Thus, $\tilde{G_1}\sync_\lor T$ is not current-state opaque. 

\item\label{item:obseq2} $x_1\in Q_1^S$ and $x_T\in Q_T^{NS}$. Then from $\tilde{G_1}\ttrans[P_G(s)][x_2]$ and $ T\ttrans[P_T(s)]x_T$, it follows that $\tilde{G}_1\sync T\ttrans[s]([x_2],x_T)$. As $x_1\in[x_2]$ based on  Definition~\ref{def:obseq}, it holds that for all $x\in [x_2]$, $x\in Q_1^S$ and $[x_2]\in \tilde{Q}_1^S$, which implies that $([x_2],x_T)$ is a secret state. Since for all $x_1'\in Q_{G_1}$ such that $G_1\ttrans[P_G(s)]x'_1$ it holds that  $x'_1\in Q_1^{S}$, based on Definition~\ref{def:obseq} it follows that for all $[x']\in \tilde{Q}_{G_1}$ such that $\tilde{G_1}\ttrans[P_G(s)] [x']$ it holds that $[x']\in \tilde{Q}^{S}_1$. Thus, $\tilde{G_1}\sync_\lor T\ttrans[s]([x'],x'_T)$ such that $[x']\in \tilde{Q}^{NS}_1$ does not exist, which means that $\tilde{G_1}\sync_\lor T$ is not current-state opaque. 

\item $x_T\in Q_T^S$ and $x_1\in Q_1^S$. Then the proof follows from \ref{item:obseq1} and \ref{item:obseq2}.
\end{enumerate}

\item Assume that $\tilde{G_1}\sync_\lor T$ is not current-state opaque. Then there exists $\tilde{G_1}\sync_\lor T\ttrans[s]([x_1],x_T)$ such that $[x_1]\in \tilde{Q}_1^S$ or $x_T\in  Q_T^S$ and there does not exists $([x'_1],x'_T)\in \tilde{Q}_{G_1}\times Q_T$ such that $\tilde{G_1}\sync_\lor T\ttrans[s]([x'_1],x'_T)$ and  $([x'_1],x'_T)\in \tilde{Q}^{NS}_1\times Q^{NS}_T$. 
From $\tilde{G_1}\sync_\lor T\ttrans[s]([x_1],x_T)$, it follows that $\tilde{G_1}\ttrans[P_G(s)][x_1]$ and $ T\ttrans[P_T(s)]x_T$. From there does not exist $([x'_1],x'_T)\in \tilde{Q}^{NS}_1\times Q^{NS}_T$ such that $\tilde{G_1}\sync_\lor T\ttrans[s]([x'_1],x'_T)$, it follows that for all $[x'_1]\in \tilde{Q}_{G_1}$ such that $\tilde{G_1}\ttrans[P_G(s)][x'_1]$ it holds that $[x'_1]\in \tilde{Q}_1^{S}$ or for all $x'_T\in Q_T$ such that $T\ttrans[P_T(s)]x'_T$ it holds that  $x'_T\in Q_T^{S}$.
Moreover, since $G_1$ and $\tilde{G_1}$ are opaque observation equivalent from  $\tilde{G_1}\ttrans[P_G(s)][x_1]$ and based on  Definition~\ref{def:obseq}, it follows that there exists $x\in [x_1]$ such that $G_1\ttrans[P_G(s)]x$.

Again, consider three cases:
\begin{enumerate}
\item\label{item:obseq3} $x_T\in Q_T^S$ and $[x_1]\in \tilde{Q}_1^{NS}$. The proof is similar to \ref{item:obseq1}.

\item \label{item:obseq4} $[x_1]\in \tilde{Q}_1^S$ and $x_T\in  Q_T^{NS}$. Then as $[x_1]\in \tilde{Q}_1^S$ it holds that for all $x\in [x_1]$ also $x\in Q_1^S$. Thus, from $G_1\ttrans[P_G(s)]x$ and $T\ttrans[P_T(s)]x_T$  it follows that there exists $G_1\sync T\ttrans[s](x,x_T)$, where $(x,x_T)$ is considered a secret state. Since for all $[x'_1]\in \tilde{Q}_{G_1}$ such that $\tilde{G_1}\ttrans[P_G(s)][x'_1]$ it holds that $[x'_1]\in \tilde{Q}_1^{S}$,
then by Definition~\ref{def:obseq} it holds that 
for all $x'\in Q_{G_1}$ such that $G_1\ttrans[P_G(s)]x'$ it holds that  $x'\in Q_1^{S}$. This
means that $G_1\sync_\lor T\ttrans[s](x',x'_T)$ such that  $x'\in Q_1^{NS}$ does not hold, which implies that $G_1\sync_\lor T$ is not current-state opaque.

\item $x_T\in Q_T^S$ and $[x_1]\in \tilde{Q}_1^S$. Then the proof follows from \ref{item:obseq3} and \ref{item:obseq4}. \QEDA
\end{enumerate}
\end{enumerate}

\Thm~\ref{thm:OpaqueOECS} illustrates that the components of a modular system that are interacting by $\sync_\lor$ can be abstracted using opaque observation equivalence while preserving the current-state opacity  property. If $\sync_\land$ is used for interaction, a similar result holds and the following corollary can be proved.
\def\OpaqueOECSLand{%
  Let $\mathcal{G}=\{G_1,\ldots, G_n\}$ be a non-deterministic system with $\sync_\land$ for interaction and with the set of secret states  $Q^S=Q_1^S\times \ldots\times Q_n^S$.  Let $\sim$ be an opaque observation equivalence on $G_1$ such that $\tilde{\SYSG}=\{\tilde{G_1},G_2,\ldots,G_n\}$. Then $\SYSG$ is current-state opaque if and only if $\tilde{\SYSG}$ is current-state opaque.
 }

\begin{corollary}\label{cor:obsEqCSLand}
\OpaqueOECSLand
\end{corollary}
\emph{Proof:}

Consider  $G_2\sync\ldots\sync G_n=T=\langle \ACT_T,Q_T,\trans,Q^\circ_T\rangle$. It suffices to show that if $G_1\sync_\land T$ is not current-state opaque then $\tilde{G}_1\sync_\land T$ is not current-state opaque either and vice versa. Let $P_G:\ACT_T\cup \ACT_{G_1}\to\ACT_G$ and $P_T:\ACT_T\cup \ACT_{G_1}\to\ACT_T$.
\begin{enumerate}
\item If $G_1\sync_\land T$ is not current-state opaque. Then there exists $(x_1,x_T)\in Q_{G_1}\times Q_T$ such that $G_1\sync_\land T\ttrans[s](x_1,x_T)$ and $(x_1,x_T)\in Q_1^S\times  Q_T^S$, and there does not exist $(x'_1,x'_T)\in Q_{G_1}\times X_T$ such that $G_1\sync_\land T\ttrans[s](x'_1,x'_T)$ and $x'_1\in Q^{NS}_1$ or $x'_T\in Q^{NS}_T$. This means, there does not exist $x'_1\in Q^{NS}_G$ or $x'_T\in Q^{NS}_T$ such that $G_1\ttrans[P_G(s)]x'_1$ or $T\ttrans[P_T(s)]x'_T$.  Now we need to show the two following cases:

\begin{enumerate}
\item $\tilde{G}_1\sync_\land T\ttrans[s]([x_2],x_T)$ and $([x_2],x_T)\in \tilde{Q}_1^S\times Q_T^{S}$. Since $G_1$ and $\tilde{G}_1$ are opaque observation equivalent from  $G_1\ttrans[P_G(s)]x_1$ and based on  Definition~\ref{def:obseq} it follows that $\tilde{G}_1\ttrans[P_G(s)][x_2]$ such that $x_1\in [x_2]$. From $\tilde{G}_1\ttrans[P_G(s)][x_2]$ and $ T\ttrans[P_T(s)]x_T$ it follows that $\tilde{G}_1\sync_\land T\ttrans[s]([x_2],x_T)$. As $x_1\in[x_2]$ based on  Definition~\ref{def:obseq} it holds that for all $x\in [x_2]$, $x\in Q_1^S$ and $[x_2]\in \tilde{Q}_1^S$. This means $([x_2],x_T)\in \tilde{Q}_1^S\times Q_T^{S}$. 

\item There does not exist $\tilde{G}_1\sync_\land T\ttrans[s]([x'],x'_T)$ such that $[x']\in \tilde{Q}_1^{NS}$ or $x'_T\in  Q_T^{NS}$. From  
there does not exist $x'_1\in Q^{NS}_G$ or $x'_T\in Q^{NS}_T$ such that $G_1\ttrans[P_G(s)]x'_1$ or $T\ttrans[P_T(s)]x'_T$
 and based on Definition~\ref{def:obseq} it follows that there does not exist $[x']\in \tilde{Q}^{NS}_1$ and $x_1'\in [x']$ such that $\tilde{G}_1\ttrans[P_G(s)] [x']$ either. This implies 
 there does not exist $\tilde{G}_1\sync_\land T\ttrans[s]([x'],x'_T)$ such that $[x']\in \tilde{Q}^{NS}_1$ or $x'_T\in Q^{NS}_T$.
 
\end{enumerate}
Therefore it can be concluded that $\tilde{G}_1\sync_\land T$ is not current-state opaque.

\item If $\tilde{G}_1\sync_\land T$ is not current-state opaque. Then there exists $\tilde{G}_1\sync_\land T\ttrans[s]([x_1],x_T)$ such that $([x_1],x_T)\in \tilde{Q}_1^S\times  Q_T^S$ and there does not exist $([x'_1],x'_T)\in \tilde{Q}_{G_1}\times Q_T$ such that $\tilde{G}_1\sync_\land T\ttrans[s]([x'_1],x'_T)$ and $[x'_1]\in \tilde{Q}^{NS}_1$ or $x'_T\in  Q^{NS}_T$. 
From $\tilde{G}_1\sync_\land T\ttrans[s]([x_1],x_T)$ it follows that $\tilde{G}_1\ttrans[P_G(s)][x_1]$ and $ T\ttrans[P_T(s)]x_T$ and $[x_1]\in\tilde{Q}_G^S$ and $x_T\in Q_T^S$. Again we need to show the two following cases:

\begin{enumerate}
\item  $G_1\sync T\ttrans[s](x,x_T)$, where $(x,x_T)\in Q^S_1\times Q^S_T$. since $G_1$ and $\tilde{G}_1$ are opaque observation equivalent from  $\tilde{G}_1\ttrans[P_G(s)][x_1]$ and based on  Definition~\ref{def:obseq} it follows that there exists $x\in [x_1]$ such that $G_1\ttrans[P_G(s)]x$. Then as $[x_1]\in \tilde{Q}_1^S$ it holds that for all $x\in [x_1]$ also $x\in Q_1^S$. Thus, from $G_1\ttrans[P_G(s)]x$ and $T\ttrans[P_T(s)]x_T$  it follows that $G_1\sync T\ttrans[s](x,x_T)$, where $(x,x_T)\in Q^S_1\times Q^S_T$.

\item $G_1\sync_\land T\ttrans[s](x',x'_T)$ such that $x'\in Q^{NS}_1$ or $x'_T\in  Q^{NS}_T$ does not exist. From there does not exist $([x'_1],x'_T)\in \tilde{Q}_{G_1}\times Q_T$ such that $\tilde{G}_1\sync_\land T\ttrans[s]([x'_1],x'_T)$ and $[x'_1]\in \tilde{Q}^{NS}_1$ or $x'_T\in  Q^{NS}_T$ and based on Definition~\ref{def:obseq} it holds that $G_1\ttrans[P_G(s)]x'$ such that $x'\in Q^{NS}_1$ does not exist. This means $G_1\sync_\land T\ttrans[s](x',x'_T)$ such that $x'\in Q^{NS}_1$ or  $x'_T\in  Q^{NS}_T$ does not exist. 
\end{enumerate}
Therefore, it can be concluded that the $G_1\sync_\land T$ is not current-state opaque.\QEDA
\end{enumerate}

\begin{example}\label{ex:oEcurrent}
Consider automaton $G_1$ in \Fig~\ref{fig:notion} with set of secret states $Q_1^S=\{s_1,s_3\}$. Consider states $s_2$ and $s_4$. Both states are non-secret states,  $s_2,s_4\in Q^{NS}$, and they have the same future behavior since $s_2\trans[\tau]s_4$ and $s_4\trans[\varepsilon]s_4$. Thus, states $s_2$ and $s_4$ can be merged while preserving opaque observation equivalence.  A similar argument holds for $s_1,s_3$. Merging $s_2$ and $s_4$, and $s_1$ and $s_3$, results in $\tilde{G_1}$ with two less states compared to $G_1$; it is shown in \Fig~\ref{fig:notion}. 
\end{example}

\subsection{Synchronous composition of CSEs}\label{sec:compos}

The idea of compositional opacity verification is to abstract the components and  construct the current-state estimator (CSE) for each abstracted component and, next, transform opacity verification to compositional nonblocking verification. 
For this approach to work, it needs to be shown that current-state opacity is preserved by synchronization of individual CSEs, step (ii) in \Fig~\ref{fig:composAlg}. 
In the following, \Propn~\ref{propos:sync} first shows that synchronization of CSEs produces an automaton  that is  isomorphic with the monolithic CSE of the original system, $det(\SYSG)$.\footnote{We could not find a proof of this result in the literature.}
Then, \Thm~\ref{thm:syncLorCO} and Corollary~\ref{cor:syncLandCO} show that the current-state opacity of a system can be verified by constructing the CSEs of individual components and then synchronizing them by using  $\sync_\lor$ or $\sync_\land$, respectively. 
As $\sync_\lor$ is more general compared with $\sync_\land$, The following proposition has been presented and proved in Chapter 5 of  \cite{SeaSilvaSch:12}, Proposition 5.5.

\begin{proposition}\label{propos:sync}\cite{SeaSilvaSch:12}
Let $G_1 = \auttuple[1]{\ACT_1}$ and $G_2 = \auttuple[2]{\ACT_2}$ be two non-deterministic automata. Then $det(G_1\sync G_2)$ is isomorphic to $det(G_1)\sync det(G_2)$.
\end{proposition}

\Propn~\ref{propos:sync} shows that the CSEs of the components of a system can be constructed  individually.\footnote{If the system components synchronize also on common \emph{unobservable} events, then \Propn~\ref{propos:sync} may not hold. } 
The idea of this paper is to transform the opacity verification problem to compositional nonblocking verification problem.  Since the  input to  compositional nonblocking verification is a set of automata, it is  essential  for the proposed algorithm to keep the modular structure of the system. 

The following theorem shows that current-state opacity is preserved by synchronization of the individual CSEs.

\def\syncLorCO{Let $\SYSG=\{G_1,\cdots,G_n\}$ be a non-deterministic system with $\sync_\lor$ for interaction, where each automaton has set of secret states $Q_i^S$. Hence, the set of secret states of the system is $Q^S=Q\setminus Q^{NS}$, where $Q^{NS}=Q^{NS}_1\times\ldots\times Q^{NS}_n$. Let $det(G_i)=\langle\ACT_i, X^i_{obs},\trans_i,X_i^\circ\rangle$ for $1\leq i\leq n$. Then $\SYSG$ is current-state opaque if and only if $det(G_1)\sync_\lor \ldots\sync_\lor det(G_n)\trans[s] (X_1,\ldots,X_n)$  implies 
 $X_i\not\subseteq Q_i^S$ for all $1\leq i\leq n$.}

\begin{theorem}\label{thm:syncLorCO}
\syncLorCO
\end{theorem}

\emph{Proof:} 

Based on \Propn~\ref{propos:sync}, it holds that $det(\SYSG)\trans[s]X$ if and only $det(G_1)\sync \ldots\sync det(G_n)\trans[s] (X_1,\ldots,X_n)$ and  $(x_1,\ldots,x_n)\in X$ if and only if $(x_1,\ldots,x_n)\in X_1\times \cdots\times X_n$. If the system is current-state opaque then  for all $det(\SYSG)\trans[s]X$ it holds that $X\not\subseteq Q^S$. Since $\sync_\lor$ is used for composition, this means there exists  $(x_1,\ldots,x_n)\in X$ such that $x_i\in Q^{NS}_i$ for all $1\leq i\leq n$. Based on \Propn~\ref{propos:sync} it holds that $det(G_1)\sync \ldots\sync det(G_n)\trans[s] (X_1,\ldots,X_n)$ and there exists $(x_1,\ldots,x_n)\in X_1\times \cdots\times X_n$ such that $x_i\in Q^{NS}_i$ for all $1\leq i\leq n$. This means that $X_i\not\subseteq Q^S_i$ for all $1\leq i\leq n$. 

Similarly, assume that $det(G_1)\sync \ldots\sync det(G_n)\trans[s] (X_1,\ldots,X_n)$ and $X_i\not\subseteq Q_i^S$ for all $1\leq i\leq n$. Since $\sync_\lor$ is used for composition, this means there exists $(x_1,\ldots,x_n)\in X_1\times\cdots\times X_n$ such that $x_i\in Q^{NS}_i$ for all $1\leq i\leq n$. Based on \Propn~\ref{propos:sync} it holds that $det(\SYSG)\trans[s]X$ and there exists $(x_1,\ldots,x_n)\in X$ such that $x_i\in Q^{NS}_i$ for all $1\leq i\leq n$, which implies $X\not\subseteq Q^S$. Thus, the system is current-state opaque. \QEDA

\def\syncLandCO{Let $\SYSG=\{G_1,\cdots,G_n\}$ be a non-deterministic system with $\sync_\land$ and with set of secret states $Q^S=Q_1^S\times\cdots\times Q_n^S$ and $det(G_i)=\langle\ACT_i, X^i_{obs},\trans_i,X_i^\circ\rangle$ for $1\leq i\leq n$. Then $\SYSG$ is current-state opaque  if and only if $det(G_1)\sync_\land \ldots\sync_\land det(G_n)\trans[s] (X_1,\cdots,X_n)$  
implies $X_1\times\cdots\times X_n\not\subseteq Q_1^S\times\cdots\times Q_n^S$.}

\Thm~\ref{thm:syncLorCO} proves that a system is current state opaque if and only if no state of the modular current-state estimator lies entirely in the set of secret states. The set of secret states in \Thm~\ref{thm:syncLorCO} is defined based on  $\sync_\lor$. The following corollary proves that a similar result also   holds when $\sync_\land$ is used for synchronization. 

\begin{corollary}\label{cor:syncLandCO}
\syncLandCO
\end{corollary}
\emph{Proof:}

Based on \Propn~\ref{propos:sync} it holds that $det(\SYSG)\trans[s]X$ if and only $det(G_1)\sync \ldots\sync det(G_n)\trans[s] (X_1,\cdots,X_n)$ and  $(x_1,\cdots,x_n)\in X$ if and only if $(x_1,\cdots,x_n)\in X_1\times\cdots\times X_n$. If the system is current-state opaque then  for all $det(\SYSG)\trans[s]X$ it holds that $X\not\subseteq Q^S$, which means that there exists $(x_1,\cdots,x_n)\in X$ such that $(x_1,\ldots,x_n)\not\in Q^S_1\times\ldots\times Q^S_n$. Based on \Propn~\ref{propos:sync} it holds that $det(G_1)\sync \ldots\sync det(G_n)\trans[s] (X_1,\cdots,X_n)$ and there exists $(x_1,\cdots,x_n)\in X_1\times\cdots\times X_n$ such that $(x_1,\ldots,x_n)\not\in Q^S_1\times\ldots\times Q^S_n$. Thus, $X_1\times\cdots\times X_n\not\subseteq  Q^S_1\times\ldots\times Q^S_n$.

Similarly assume $det(G_1)\sync \ldots\sync det(G_n)\trans[s] (X_1,\cdots,X_n)$ and  $X_1\times\cdots\times X_n\not\subseteq  Q^S_1\times\ldots\times Q^S_n$. This means that there exists  $(x_1,\ldots,x_n)\in X_1\times\cdots\times X_n$ such that $(x_1,\ldots,x_n)\not\in Q^S_1\times\ldots\times Q^S_n$.  Based on \Propn~\ref{propos:sync} it holds that $det(\SYSG)\trans[s]X$ and there exists $(x_1,\cdots,x_n)\in X$ such that $(x_1,\ldots,x_n)\not\in Q^S_1\times\ldots\times Q^S_n$, which implies $X\not\subset Q^S$. Thus, the system is current-state opaque. \QEDA

\begin{figure}
	\centering
		\includegraphics{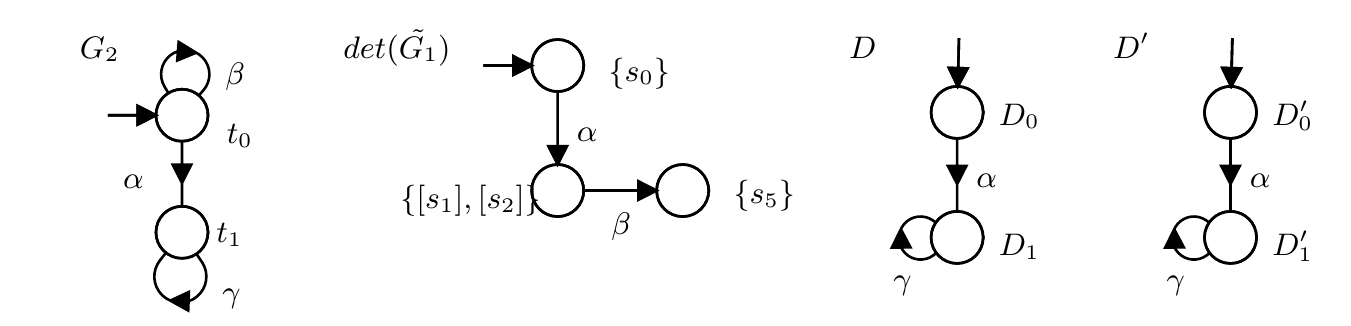}\caption{Automata of \Examp~\ref{ex:syncCurrent}.}\label{fig:syncIso}
\end{figure}

\begin{example}\label{ex:syncCurrent}
Assume we have the system $\SYSG=\{\tilde{G}_1,G_2\}$, where the set of secret states of $\tilde{G}_1$ is $\tilde{Q}^S_1=\{[s_1],[s_2]\}$ and $Q^S_2=\{t_1\}$. Automaton $\tilde{G}_1$ and $G_2$ are shown in \Fig~\ref{fig:notion} and \Fig~\ref{fig:syncIso}, respectively. \Fig~\ref{fig:syncIso} also shows $D=det(\tilde{G}_1\sync G_2)$ and $D'=det(\tilde{G}_1)\sync det(G_2)$, where 
\begin{itemize}
\item $D_0=\{(s_0,t_0)\}$, $D_1=\{([s_1],t_1),([s_2],t_1)\}$ in $D$
\item $D'_0=(\{s_0\},\{t_0\})$, $D'_1=(\{[s_1],[s_2]\},\{t_1\})$ in $D'$.
\end{itemize}
It can be observed that $D$ and $D'$ are isomorphic. If $\sync_\lor$ is used for synchronization, then $([s_1],t_1)$ and $([s_2],t_1)$ are considered secret states. This means that $D_1\subseteq Q^S$. Therefore, it can be concluded that $G_1\sync_\lor G_2$ is not current-state opaque. Moreover, $([s_1],t_1)$ and $([s_2],t_1)$ are also considered secret if $\sync_\land$ is used for interaction. Therefore, $G_1\sync_\land G_2$ is not current-state opaque.
\end{example}

\subsection{Current-state opacity to nonblocking verification}\label{sec:currentNon}

So far, we have shown that the components of a modular system can be abstracted using opaque observation equivalence and the CSEs of individual abstracted components can be built.  
This section describes the transformation of compositional opacity verification to nonblocking verification, steps (iii) and (iv) in \Fig~\ref{fig:composAlg}. 
These steps are done after  creating the modular current-state estimators. 
The compositional approach is well-established for nonblocking verification \cite{FloMal:09,WarMal:10}. 
A variety of abstraction methods with efficient implementations that preserve the  nonblocking property are introduced in  \cite{FloMal:09,WarMal:10}. 
Thus, transforming current-state opacity to nonblocking verification makes it possible to use well-developed algorithms for nonblocking verification. 
To transform current-state opacity to nonblocking verification, the first step is to identify the states of individual current-state estimators that are violating opacity. 
These states are considered as $\psi$-states in \Defn~\ref{def:omegaAutomaton}. 
From these states, transitions to blocking states are added. 
The system is current-state opaque if and only if the transformed system is nonblocking. 


As explained in \Sect~\ref{sec:preli}, the components of the system can interact either by $\sync_\lor$ or $\sync_\land$. Therefore, two different transformations are required to reflect  this. In the following, \Thm~\ref{them:OpaqueToBlockigCSLor} and Corollary~\ref{them:OpaqueToBlockigCSLand} formally describe how to transform current-state opacity to nonblocking verification when $\sync_\lor$ and $\sync_\land$ are used, respectively. 

\def\OpaqueToBlockigCSLor{Let $\mathcal{G}=\{G_1,\cdots,G_n\}$ be a non-deterministic system, with $\sync_\lor$ used for interaction, where each automaton has set of secret states $Q_i^S$. Hence, the set of secret states of the system is $Q^S=Q\setminus Q^{NS}$, where $Q^{NS}=Q^{NS}_1\times\ldots\times Q^{NS}_n$. Let $det(G_i)=\langle \ACT_i, X_{obs}^i,\trans_{i},X_i^\circ\rangle$. Then $\SYSG$ is not current-state opaque if and only if $det(G_1)_{\psi_1}\sync \ldots \sync det(G_n)_{\psi_n}$ is blocking, where $X_{\psi_i}^i = \{X\in X_{obs}^i\ |\ X\subseteq Q_i^S\}$.}

\begin{theorem}\label{them:OpaqueToBlockigCSLor}
\OpaqueToBlockigCSLor
\end{theorem}

\emph{Proof:} 

First, assume that $\SYSG$ is not current-state opaque. Based on \Thm~\ref{thm:syncLorCO} it holds that $det(G_1)\sync\ldots\sync det(G_n)\trans[s](X_1,\cdots,X_n)$ such that  $X_i\subseteq Q_i^S$ for some $1\leq i\leq n$, which implies $X_i\in X_{\psi_i}^i$. Based on \Lemm~\ref{lemm:blockingForLor} in Appendix, $det(G_1)_{\psi_1}\sync \ldots \sync det(G_n)_{\psi_n}$ is blocking. 

Next, assume that $det(G_1)_{\psi_1}\sync \ldots \sync det(G_n)_{\psi_n}$ is blocking. Then based on \Lemm~\ref{lemm:blockingForLor} in Appendix, it holds that $det(G_1)\sync\ldots\sync det(G_n)\trans[s](X_1,\cdots,X_n)$ such that $X_i\in X_{obs}^i$ and $X_i\in X_{\psi_i}^i$ for some $1\leq i\leq n$. This means that $det(G_1)\sync\ldots\sync det(G_n)\trans[s](X_1,\cdots,X_n)$ such that $X_i\subseteq Q_i^S$ for some $1\leq i\leq n$, which based on \Thm~\ref{thm:syncLorCO} implies that $\SYSG$ is not current-state opaque. \QEDA

\begin{figure}
	\centering
		\includegraphics{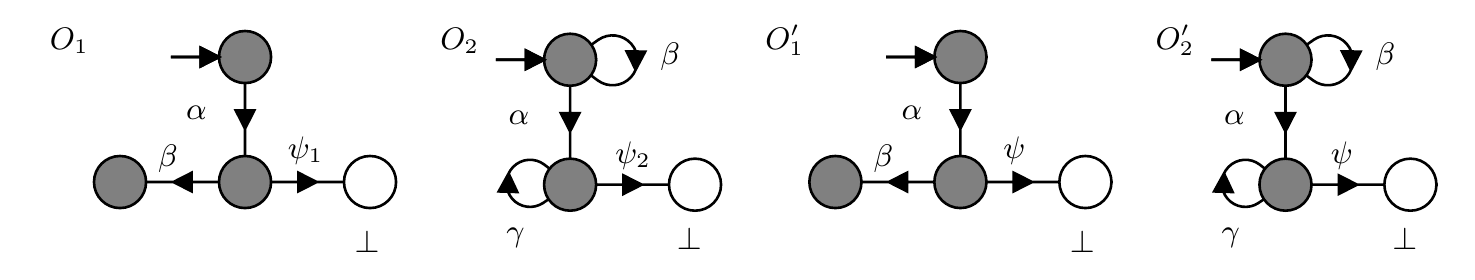}
\caption{Automata of \Examp~\ref{ex:nonCurrentLor} and~\ref{ex:nonCurrentLand}.}\label{fig:nonCurrent}
\end{figure}

The following example shows all the steps of compositional current-state opacity verification. 
\begin{example}\label{ex:nonCurrentLor}
Consider the system $\SYSG=\{G_1,G_2\}$, where $G_1$ is shown in \Fig~\ref{fig:notion} and $G_2$ is shown in \Fig~\ref{fig:syncIso}. Assume that the set of secret states of $G_1$ and $G_2$ are $Q^S_1=\{s_1,s_2,s_3,s_4\}$ and $Q_2^S=\{t_1\}$, respectively. At the first step of the compositional approach, step (i) in Fig~\ref{fig:composAlg}, each automaton is abstracted using opaque observation equivalence. Thus, $G_1$ is replaced by $\tilde{G}_1$, shown in \Fig~\ref{fig:notion}, and $G_2$ can not be abstracted. Automaton $\tilde{G}_1$ has set of secret states $\tilde{Q}^S_1=\{[s_1],[s_2]\}$. Next, $det(\tilde{G}_1)$ and $det(G_2)$ are constructed, step (ii) in \Fig~\ref{fig:composAlg}. \Fig~\ref{fig:syncIso} shows $det(\tilde{G}_1)$ and since $G_2$ is deterministic, $det(G_2)=G_2$. The next step in the compositional approach is to transform  opacity verification to nonblocking verification by building the $\psi$-automata of the individual components, step (iii) in \Fig~\ref{fig:composAlg}. Assume that $\sync_\lor$ is used for interaction. As $\{[s_1],[s_2]\}\subseteq Q^S_1$ in $det(\tilde{G_1})$ and $\{t_1\}\subseteq Q^S_2$ in $det(G_2)$, it follows that $X_{\psi_1}^1=\{\{[s_1],[s_2]\}\}$ and $X_{\psi_2}^2=\{\{t_1\}\}$. The $\psi$-automata of $det(\tilde{G}_1)$ and $det(G_2)$ are shown in \Fig~\ref{fig:nonCurrent} as $O_1$ and $O_2$, respectively. The transformed system is blocking as $O_1\sync O_2\trans[\alpha\psi_1]\dumpstate$, step (iv) in \Fig~\ref{fig:composAlg}. Thus, we can conclude that the original system is not current-state opaque when $\sync_\lor$ is used for interaction. Indeed, the system is not current-state opaque as if $\alpha$ occurs in the monolithic system,  then $G_1\sync_\lor G_2$ goes to the secret states $(s_1,t_1)$ or $(s_2,t_1)$, from where there is no possible denial. 
\end{example}

\def\OpaqueToBlockigCSLand{Let $\mathcal{G}=\{G_1,\cdots,G_n\}$ be a non-deterministic system, with $\sync_\land$ used for interaction, where the set of secret states is  $Q^S=Q_1^S\times\cdots\times Q_n^S$. Let $det(G_i)=\langle \ACT_i, X_{obs}^i,\trans_{i},X_i^\circ\rangle$. Then $\SYSG$ is not current-state opaque if and only if $det(G_1)_{\psi}\sync \ldots \sync det(G_n)_{\psi}$ is blocking, where $X_\psi^i = \{X\in X_{obs}^i\ |\ X\subseteq Q_i^S\}$. }

\begin{corollary}\label{them:OpaqueToBlockigCSLand}
\OpaqueToBlockigCSLand
\end{corollary}
\emph{Proof:}

First assume $\SYSG$ is not current-state opaque. Based on Definition~\ref{def:currentStateOp} it holds that there exists $s\in \ACT^*$ such that $det(\SYSG)\trans[s]X$ and  $X\subseteq Q^S$. Based on \Propn~\ref{propos:sync} $det(\SYSG)$ is isomorphic to $det(G_1)\sync \ldots \sync det(G_n)$. Thus, $det(G_1)\sync\ldots\sync det(G_n)\trans[s](X_1,\cdots,X_n)$ and $X_1\times\cdots\times X_n\subseteq Q_1^S\times\ldots\times Q_n^S$, which implies  $X_i\subseteq Q_i^S$ for all $1\leq i\leq n$. Thus, it holds that  $X_1\times \cdots\times X_n\subseteq X_\psi^1\times\cdots\times X_\psi^n$. Based on Lemma~\ref{lemm:blockingForAnd}, $det(G_1)_\psi\sync \ldots \sync det(G_n)_\psi$ is blocking. 

Now assume $det(G_1)_{\psi}\sync \ldots \sync det(G_n)_{\psi}$ is blocking. Then based on Lemma~\ref{lemm:blockingForAnd} it holds that $det(G_1)\sync\ldots\sync det(G_n)\trans[s](X_1,\cdots,X_n)$ and  $X_1\times\cdots\times X_n\subseteq X_\psi^1\times\cdots\times X_\psi^n$. This means $det(G_1)\sync\ldots\sync det(G_n)\trans[s](X_1,\cdots,X_n)$ and $X_i\subseteq Q_i^S$ for all $1\leq i\leq n$, which implies that $\SYSG$ is not current-state opaque. \QEDA

The main difference between \Thm~\ref{them:OpaqueToBlockigCSLor} and Corollary~\ref{them:OpaqueToBlockigCSLand} is in generating the $\psi$-automata of the individual components. When $\sync_\lor$ is used for composition, then the system is not current-state opaque if at least one state of the composed states is secret. To capture this feature, individual components $G_i$ have different $\psi_i$-transitions. This guarantees that if the transformed system is blocking, then there is a composed state in the original system with at least one secret state. In contrast, when $\sync_\land$ is used for interaction, a composed state is considered secret if all the states are secret. To assure this, in the transformed system all the components have the same $\psi$-transitions. Thus, the $\psi$-transitions  happen if and only if they happen in all the components simultaneously.

\begin{example}\label{ex:nonCurrentLand}
Consider again the system $\SYSG=\{\tilde{G}_1,G_2\}$, where $\tilde{G}_1$ is shown in \Fig~\ref{fig:notion} and $G_2$ is shown in \Fig~\ref{fig:syncIso}. Assume that the set of secret states of $\tilde{G}_1$ and $G_2$ are $\tilde{Q}^S_1=\{[s_1],[s_2]\}$ and $Q_2^S=\{t_1\}$, respectively. Now, assume that $\sync_\land$ is used for interaction. Similarly to \Examp~\ref{ex:nonCurrentLor}, we have that $X_{\psi}^1=\{\{[s_1],[s_2]\}\}$ and $X_{\psi}^2=\{\{t_1\}\}$. In contrast to \Examp~\ref{ex:nonCurrentLor}, to generate the $\psi$-automata when $\sync_\land$ is used for interaction, all the transitions to $\dumpstate$ are labeled by the same $\psi$ event. The $\psi$-automata of $det(\tilde{G}_1)$ and $det(G_2)$ are  shown in \Fig~\ref{fig:nonCurrent} as $O'_1$ and $O'_2$, respectively. The $\psi$-system is blocking as $O'_1\sync O'_2\trans[\alpha\psi]\dumpstate$. Thus, we can conclude that the original system is not current-state opaque when $\sync_\land$ is used for interaction. 
\end{example}


\section{Compositional infinite-step opacity verification}\label{sec:Compoinf}

This section discusses the compositional approach for infinite-step opacity verification. In the compositional infinite-step   opacity verification individual components are first abstracted. Next, the two-way observers \cite{XiangLaf:17} of components are  generated and finally the infinite-step opacity verification is transformed to the compositional nonblocking verification. In the following, Section~\ref{sec:ooeInf} explains how individual components can be abstracted before building the modular two-way observers. Next, Section~\ref{sec:syncInf} establishes a link between the modular two-way observers and the monolithic two-way observer. Section~\ref{sec:infNon} shows the transformation of the infinite-step opacity verification to nonblocking verification when $\sync_\lor$ and $\sync_\land$ is used for interaction.

\subsection{Opaque observation equivalence}\label{sec:ooeInf}

As explained in Section~\ref{sec:compo}, the first step of the compositional infinite-step opacity  is to abstract the components using opaque observation equivalence. In the following Theorem~\ref{them:obseqInfinit} and Corollary~\ref{cor:obseqInfinit} show that opaque observation equivalence preserves infinite-step opacity when $\sync_\lor$ and $\sync_\land$ are used for interaction, respectively. Theorem~\ref{them:obseqInfinit} and Corollary~\ref{cor:obseqInfinit} are similar to  Theorem~\ref{thm:OpaqueOECS} and Corollary~\ref{cor:obsEqCSLand} presented 
in Section~\ref{sec:obsEqCurrent}, and accordingly, they have similar proofs.

\def\oseqInfini{  Let $\mathcal{G}=\{G_1,\ldots, G_n\}$ be a non-deterministic system with $\sync_\lor$ for interaction, where each automaton has  set of secret states $Q_i^S$. Hence, the set of secret states of the system is $Q^S=Q\setminus Q^{NS}$, where $Q^{NS}=Q^{NS}_1\times\ldots\times Q^{NS}_n$.  Let $\sim$ be an opaque observation equivalence on $G_1$ such that $\tilde{\SYSG}=\{\tilde{G_1},G_2,\ldots,G_n\}$. Then $\SYSG$ is infinite-step opaque if and only if $\tilde{\SYSG}$ is infinite-step opaque.}

\begin{theorem}\label{them:obseqInfinit}
\oseqInfini
\end{theorem}

\emph{Proof:} 

Assume $T= G_2\sync_\lor\ldots \sync_\lor G_n$ with  set of secret states $Q^S_T$ and set of non secret states $Q_T^{NS}=Q_T\setminus Q_T^S$. Then $Q^S_T=Q^T\setminus Q_T^{NS}$, where $Q_T^{NS}=Q^{NS}_2\times\ldots\times Q^{NS}_n$. It suffices to show that if $G_1\sync_\lor T$ is not infinite-step opaque then $\tilde{G}_1\sync_\lor T$ is not infinite-step opaque either and vice versa. 
\begin{enumerate}
\item First assume that $G_1\sync_\lor T$ is not infinite-step opaque. Then there exists $(x^0_G,x^0_T)\in Q^\circ_{G_1}\times Q^\circ_T$ and $st\in\LANG({G_1}\sync_\lor T,(x^0_G,x^0_T))$ such that $(x^0_G,x^0_T)\ttrans[s](x_1,x_T)$, and $x_1\in Q_{G_1}^S$ or $x_T\in  Q_T^S$, and  one of the following holds.
\begin{itemize}
\item For all $(x'_1,x'_T)\in Q^{NS}_{G_1}\times Q^{NS}_T$ such that ${G_1}\sync_\lor T\ttrans[s](x'_1,x'_T)$ then ${G_1}\sync_\lor T\ttrans[s](x'_1,x'_T)\not\ttrans[t]$. 
\item There does not exist $(x'_1,x'_T)\in Q^{NS}_{G_1}\times Q^{NS}_T$ such that ${G_1}\sync_\lor T\ttrans[s](x'_1,x'_T)$. 

\end{itemize} 
Since ${G_1}\sim \tilde{G}_1$ it holds that $\tilde{G}_1\sync_\lor T \sim {G_1}\sync_\lor T$. This means, if ${G_1}\sync_\lor T\ttrans[s](x'_1,x'_T)\not\ttrans[t]$ then $\tilde{G}_1\sync_\lor T\ttrans[s]([x'_1],x'_T)\not\ttrans[t]$, where  $x'_1\in[x_1']$ and $(x'_1,x'_T)\in Q^{NS}_{G_1}\times Q^{NS}_T$ if and only if $([x'_1],x'_T)\in \tilde{Q}^{NS}_{G_1}\times Q^{NS}_T$. Thus, $\tilde{G}_1\sync_\lor T$ is not infinite-step opaque either. 

 Now consider the case when $G_1\sync_\lor T$ is not infinite-step opaque because $(x^0_G,x^0_T)\ttrans[s](x_1,x_T)$, and $x_1\in Q_{G_1}^S$ or $x_T\in  Q_T^S$ and there does not exist $(x'_1,x'_T)\in Q^{NS}_{G_1}\times Q^{NS}_T$ such that ${G_1}\sync_\lor T\ttrans[s](x'_1,x'_T)$.
The proof for this part is the same as the proof for Theorem~\ref{thm:OpaqueOECS} in Section~\ref{sec:obsEqCurrent}, proof of opaque observation equivalence preserves current-state opacity, and consequently omitted.

\item If $\tilde{G}_1\sync_\lor T$ is not infinite-step opaque. Then there exists $([x_G^0],x_T^0)\in \tilde{Q}^\circ_{G_1}\times Q_T^\circ$ and $st\in\LANG(\tilde{G}_1\sync_\lor T,([x^0_G],x^0_T))$ such that $([x^0_G],x^0_T)\ttrans[s]([x_1],x_T)$, and $[x_1]\in \tilde{Q}_{G_1}^S$ or $x_T\in  Q_T^S$, and one of the following cases holds.
\begin{itemize}
\item For all $([x'_1],x'_T)\in \tilde{Q}^{NS}_{G_1}\times Q^{NS}_T$ such that $\tilde{G}_1\sync_\lor T\ttrans[s]([x'_1],x'_T)$ then $\tilde{G}\sync_\lor T\ttrans[s]([x'_1],x'_T)\not\ttrans[t]$.
\item There does not exist $([x'_1],x'_T)\in \tilde{Q}^{NS}_{G_1}\times Q^{NS}_T$ such that $\tilde{G}_1\sync_\lor T\ttrans[s]([x'_1],x'_T)$. 
\end{itemize} 
Since ${G_1}\sim \tilde{G}_1$ it holds that $\tilde{G}_1\sync_\lor T \sim G\sync_\lor T$. This means, if $\tilde{G}_1\sync_\lor T\ttrans[s]([x'_1],x'_T)\not\ttrans[t]$ then ${G_1}\sync_\lor T\ttrans[s](x'_1,x'_T)\not\ttrans[t]$, where  $x'_1\in[x_1']$ and $(x'_1,x'_T)\in Q^{NS}_{G_1}\times Q^{NS}_T$ if and only if $([x'_1],x'_T)\in \tilde{Q}^{NS}_{G_1}\times Q^{NS}_T$. Thus, ${G_1}\sync_\lor T$ is not infinite-step opaque either.   

Now consider the case when $\tilde{G}_1\sync_\lor T$ is not infinite-step opaque because $([x^0_G],x^0_T)\ttrans[s]([x_1],x_T)$, and $[x_1]\in \tilde{Q}_{G_1}^S$ or $x_T\in  Q_T^S$ and there does not exist $([x'_1],x'_T)\in \tilde{Q}^{NS}_{G_1}\times Q^{NS}_T$ such that $\tilde{G}_1\sync_\lor T\ttrans[s]([x'_1],x'_T)$.
Again the proof for this part is the same as the proof for Theorem~\ref{thm:OpaqueOECS} in Section~\ref{sec:obsEqCurrent}, proof of opaque observation equivalence preserves current-state opacity, and consequently omitted.\QEDA 

\end{enumerate}

In the theorem $\sim_{inf}$ refers to infinite-step opaque equivalence definite in Section~\ref{sec:Compocur}.
\def\oseqInfiniland{ Let $\mathcal{G}=\{G_1,\ldots, G_n\}$ be a non-deterministic system with $\sync_\land$ for interaction and with the set of secret states  $Q^S=Q_1^S\times \ldots\times Q_n^S$.  Let $\sim$ be an opaque observation equivalence on $G_1$ such that $\tilde{\SYSG}=\{\tilde{G_1},G_2,\ldots,G_n\}$. Then $\SYSG$ is infinite-step opaque if and only if $\tilde{\SYSG}$ is infinite-step opaque.}

\begin{corollary}\label{cor:obseqInfinit}
\oseqInfiniland
\end{corollary}

\emph{Proof}: 

Assume $T= G_2\sync_\land\ldots \sync_\land G_n$ with  set of secret states $Q^S_T$ and set of non secret states $Q_T^{NS}=Q_T\setminus Q_T^S$. Then $Q^S_T=Q^T\setminus Q_T^{NS}$, where $Q_T^{S}=Q^{S}_2\times\ldots\times Q^{S}_n$. It suffices to show that if $G_1\sync_\land T$ is not infinite-step opaque then $\tilde{G}_1\sync_\land T$ is not infinite-step opaque either and vice versa. 
\begin{enumerate}
\item  First assume that $G_1\sync_\land T$ is not infinite-step opaque. Then there exists $(x^0_G,x^0_T)\in Q^\circ_{G_1}\times Q^\circ_T$ and $st\in\LANG(G_1\sync_\land T,(x^0_G,x^0_T))$ such that $(x^0_G,x^0_T)\ttrans[s](x_1,x_T)$, and $(x_1,x_T)\in Q_{G_1}^S\times   Q_T^S$, and  one of the following holds.
\begin{itemize}
\item For all $(x'_1,x'_T)$ such that $x'_1\in Q^{NS}_{G_1}$ or $x'_T\in Q^{NS}_T$ and ${G_1}\sync_\land T\ttrans[s](x'_1,x'_T)$ then ${G_1}\sync_\land T\ttrans[s](x'_1,x'_T)\not\ttrans[t]$. 
\item There does not exist $x'_1\in Q^{NS}_{G_1}$ or $x'_T\in Q^{NS}_T$ such that ${G_1}\sync_\land T\ttrans[s](x'_1,x'_T)$.

\end{itemize} 
Since ${G_1}\sim \tilde{G}_1$ it holds that $\tilde{G}_1\sync_\land T \sim G_1\sync_\land T$. This means, if $G_1\sync_\land T\ttrans[s](x'_1,x'_T)\not\ttrans[t]$ then $\tilde{G}_1\sync_\land T\ttrans[s]([x'_1],x'_T)\not\ttrans[t]$, where  $x'_1\in[x_1']$ and $x'_1\in Q^{NS}_{G_1}$ if and only if $[x'_1]\in \tilde{Q}^{NS}_{G_1}$. Thus, $\tilde{G}_1\sync_\land T$ is not infinite-step opaque either.

 Now consider the case, where $(x_1,x_T)\in Q_{G_1}\times Q_T$ such that ${G_1}\sync_\land T\ttrans[s](x_1,x_T)$ and $(x_1,x_T)\in Q_{G_1}^S\times  Q_T^S$, and there does not exist $x'_1\in Q^{NS}_{G_1}$ or $x'_T\in Q^{NS}_T$ such that ${G_1}\sync_\land T\ttrans[s](x'_1,x'_T)$. The proof for this part is the same as the proof for Corollary~\ref{cor:obsEqCSLand}, proof of opaque observation equivalence preserves current-state opacity, and consequently omitted. 

Therefore, it can be concluded that $\tilde{G}_1\sync_\land T$ is not infinite-step opaque.

\item If $\tilde{G}_1\sync_\land T$ is not infinite-step opaque. Then there exists $([x_G^0],x_T^0)\in \tilde{Q}^\circ_{G_1}\times Q_T^\circ$ and $st\in\LANG(\tilde{G}_1\sync_\land T,([x^0_G],x^0_T))$ such that $([x^0_G],x^0_T)\ttrans[s]([x_1],x_T)$, and  $([x_1],x_T)\in \tilde{Q}_{G_1}^S\times  Q_T^S$, and one of the following cases holds.
\begin{itemize}
\item For all $([x'_1],x'_T)\in \tilde{Q}_{G_1}\times Q_T$  such that $[x'_1]\in \tilde{Q}^{NS}$ or $x'_T\in  Q^{NS}_T$ and $\tilde{G}_1\sync_\land T\ttrans[s]([x'_1],x'_T)$ then $\tilde{G}_1\sync_\land T\ttrans[s]([x'_1],x'_T)\not\ttrans[t]$.
\item There does not exist $([x'_1],x'_T)\in \tilde{Q}^{NS}_{G_1}\times Q^{NS}_T$ such that $\tilde{G}_1\sync_\land T\ttrans[s]([x'_1],x'_T)$. 
\end{itemize} 
Since $G\sim \tilde{G}_1$ it holds that $\tilde{G}_1\sync_\land T \sim G_1\sync_\land T$. This means, if $\tilde{G}_1\sync_\land T\ttrans[s]([x'_1],x'_T)\not\ttrans[t]$ then $G_1\sync_\land T\ttrans[s](x'_1,x'_T)\not\ttrans[t]$, where  $x'_1\in[x_1']$ and $x'_1\in Q^{NS}_{G_1}$ if and only if $[x'_1]\in \tilde{Q}^{NS}_{G_1}\times Q^{NS}$. Thus, $G_1\sync_\land T$ is not infinite-step opaque either.

 Now consider the case where, $\tilde{G}_1\sync_\land T$ is not infinite-step opaque because there exists $\tilde{G}_1\sync_\land T\ttrans[s]([x_1],x_T)$ such that $([x_1],x_T)\in \tilde{Q}_{G_1}^S\times  Q_T^S$ and $\tilde{G}_1\sync_\land T\not\ttrans[s]([x'_1],x'_T)$ such that $[x'_1]\in \tilde{Q}^{NS}_{G_1}$ or $x'_T\in  Q^{NS}_T$. 
Again the proof for this part is the same as the proof for Corollary~\ref{cor:obsEqCSLand}, proof of opaque observation equivalence preserves current-state opacity, and consequently omitted. 

Therefore, it can be concluded that the $G\sync_\land T$ is not infinite-step opaque.\QEDA
\end{enumerate}

\subsection{Synchronous composition of two-way observers}\label{sec:syncInf}
It has been shown in \cite{XiangLaf:17} that a two-way observer can be used to  verify infinite-step opacity. In order to monolithically verify the infinite-step opacity of a modular system the monolithic representation of whole system is generated first. Next, the two-way observer of the system is constructed, see Definition~\ref{def:twObs}. This approach requires generating two observer automata for two potentially large components, which maybe intractable. The compositional approach on the other hand, avoids constructing the monolithic two-way observer of the system. Instead, the two-way observers of the components are built individually. However, the monolithic two-way observer of the system is a subautomaton of the composed individual two-way observers. This is shown in Proposition~\ref{pro:reverseNotisomorphic}.

\begin{proposition}\label{pro:reverseNotisomorphic}
Let $G_1$ and $G_2$ be two non-deterministic automata. Let $G_{i,R}$ be the reversed automaton of $G_i$ for $i=1,2$  and let $G_R$ be the reversed automaton of $G_1\sync G_2$. Let $\Delta\colon \ACT\trans\ACT\times\{\epsilon\}$ and $\Delta_R\colon \ACT\trans\{\epsilon\}\times\ACT$. Let $H_i=\Delta(det(G_i))\sync \Delta_R(det(G_{i,R}))$ for $i=1,2$ and $H=\Delta(det(G_1\sync G_2))\sync\Delta_R(det(G_R))$. Then $H\sqsubseteq	  H_1\sync H_2$, which means that if $H\trans[s](X,X_R)\trans[\sigma] (Y,Y_R)$ in $H$ then $H_1\sync H_2\trans[s]((X^1,X^1_R),(X^2,X^2_R))\trans[\sigma] ((Y^1,Y^1_R),(Y^2,Y^2_R))$ in $H_1\sync H_2$ such that
$X_R\subset X^1_R\times X^2_R$ and 
$Y_R\subseteq Y^1_R\times Y^2_R$.
\end{proposition}

\emph{Proof:}

Based on \Lemm~\ref{lemm:renamingInsync} in Appendix, it holds that $\Delta(det(G_1))\sync \Delta(det(G_2)) = \Delta(det(G_1)\sync det(G_2))$, and based on \Propn~\ref{propos:sync},  $\Delta(det(G_1)\sync det(G_2))$ and $\Delta(det(G_1\sync G_2))$ are isomorphic. Thus, it is enough to show that $\Delta_R(det(G_R))\sqsubseteq	 \Delta_R(det(G_{1,R}))\sync\Delta_R(det(G_{2,R}))$. Moreover, based on \Lemm~\ref{lemm:renamingInsync}, it holds that $\Delta_R(det(G_{1,R}))\sync\Delta_R(det(G_{2,R}))=\Delta_R(det(G_{1,R})\sync det(G_{2,R}))$. Thus, it is enough to show that $det(G_R)\sqsubseteq	  det(G_{1,R})\sync det(G_{2,R})$.

 It is shown by induction on $n\geq 0$ that $X^0\trans[\sigma_1]X^1\trans[\sigma_2]\ldots\trans[\sigma_n]X_n$ in  $det(G_R)$ if $(X^0_1,X^0_2)\trans[\sigma_1](X_1^1,X_2^1)\trans[\sigma_2]\ldots\trans[\sigma_n](X_1^n,X_2^n)$ in $det(G_{1,R})\sync det(G_{2,R})$ such that $(x_1,x_2)\in X^j$ if $(x_1,x_2)\in X^j_1\times X^j_2$.

\emph{Base case:} $n=0$. Let $X_{R}^\circ$ be the initial state of $det(G_R)$ and $X_{i,R}^\circ$ be the initial state of $det(G_{i,R})$ for $i=1,2$. If $(x_1,x_2)\in X_{R}^\circ$, then as all the states of $G_R$ are considered as initial states it holds that $(x_1,x_2)$ is a  reachable state, which means that $G_1\sync G_2\ttrans[s](x_1,x_2)$. Based on \Defn~\ref{def:synch}, it holds that $G_i\ttrans[P_i(s)]x_i$ in $G_i$ for $i=1,2$. Based on the definition of the reversed automaton, all the states of $G_{i,R}$ for $i=1,2$ are considered as initial states, which implies that $x_i\in Q^\circ_{{i,R}}$  for $i=1,2$, . Thus, $(x_1,x_2)\in Q^\circ_{{1,R}}\times Q^\circ_{{2,R}}$, which means that $(x_1,x_2)\in X_{1,R}^\circ\times X_{2,R}^\circ$.

\emph{Inductive step:} Assume that the claim holds for some $n\geq 0$, i.e,  $X_R^\circ=X^0\trans[\sigma_1\sigma_2\ldots\sigma_n]X^n=X$ in $det(G_R)$ if $(X_{1,R}^\circ,X_{2,R}^\circ)=(X_1^0,X_2^0)\trans[\sigma_1\sigma_2\ldots\sigma_n](X^n_1,X^n_2)=(X_1,X_2)$ in $det(G_{1,R})\sync det(G_{2,R})$, and  $(x_1,x_2)\in X^k$ if  $(x_1,x_2)\in X^k_1\times X^k_2$ for all $0\leq k< n$. It must be shown that if $X=X^n\trans[\sigma_{n+1}]Y$ in $det(G_R)$, then $(X_1,X_2)=(X^n_1,X^n_2)\trans[\sigma_{n+1}](Y_1,Y_2)$  in $det(G_{1,R})\sync det(G_{2,R})$,  and if $(x_1,x_2)\in X$, then $(x_1,x_2)\in X_1\times X_2$. 

Let $det(G_R)\trans[\sigma_1\ldots\sigma_n]X^n =X\trans[\sigma_{n+1}] Y$ in $det(G_R)$  and $(x_1,x_2)\in X$. Then based on $UR(x)$ it follows that  $(x_1,x_2)=(x_1^1,x_2^1)\ttrans[\tau](x_1^r,x_2^t)\trans[\sigma_{n+1}](y_1,y_2)$ in $G_R$. Since $(y_1,y_2)$ is a reachable state in $G_R$, it holds that there exists $s\in(\ACT_1\cup\ACT_2)^*$ such that $G_1\sync G_2\ttrans[s](y_1,y_2)\trans[\sigma_{n+1}](x_1^r,x_2^t)\ttrans[\tau](x_1^1,x_2^1)$. Based on \Defn~\ref{def:synch}, it holds that $G_1\ttrans[P_1(s)]y_1\trans[P_1(\sigma_{n+1})]x_1^r\ttrans[\tau]x_1^1$ in $G_1$ and $G_2\ttrans[P_2(s)]y_2\trans[P_2(\sigma_{n+1})]x_2^t\ttrans[\tau]x_2^1$ in $G_2$. This means that $x_1^1\ttrans[\tau]x_1^r\trans[P_1(\sigma_{n+1})]y_1$ in $G_{1,R}$ and $x_2^1\ttrans[\tau]x_2^t\trans[P_2(\sigma_{n+1})]y_2$ in $G_{2,R}$. Then based on $UR(x)$ and the inductive hypothesis, it holds that $det(G_{1,R})\trans[P_1(\sigma_1\sigma_2\ldots\sigma_n)]X_1\trans[P_1(\sigma_{n+1})]Y_1$ and $det(G_{2,R})\trans[P_2(\sigma_1\sigma_2\ldots\sigma_n)]X_2\trans[P_2(\sigma_{n+1})]Y_2$. Moreover, based on $UR(x)$ it holds that   $x_1\in X_1$ and $x_2\in X_2$. Then based on \Defn~\ref{def:synch} it holds that $det(G_{1,R})\sync det(G_{2,R})\trans[\sigma_1\sigma_2\ldots\sigma_n](X_1,X_2)\trans[\sigma_{n+1}](Y_1,Y_2)$. Thus, $(X_1,X_2)\trans[\sigma_{n+1}](Y_1,Y_2)$ in $det(G_{1,R})\sync det(G_{2,R})$ and $(x_1,x_2)\in X_1\times X_2$.\QEDA

Proposition~\ref{pro:reverseNotisomorphic}  only establishes that the monolithic two-way observer of a modular system is a subautomaton of the synchronous product of the  individual two-way observers but in contrast to Proposition~\ref{propos:sync} the isomorphism does not hold in general. The reason is when the monolithic representation of the system is built some combinations of states become unreachable, or some transitions become disabled. However, those unreachable states may become reachable when individual reverse automata are composed. This happens because in the reversed automaton all the states are considered as initial state and consequently reachable states. This over approximation may cause some unreachable states that are violating the infinite-step opacity to  become reachable. Therefore, the verification results may return incorrect violation of infinite-step opacity. This is shown in Example~\ref{ex:twowayInfiniLor}.

\def\synctwoInfinit{Let $\mathcal{G}=\{G_1,\ldots, G_n\}$ be a nondeterministic system with $\sync_\lor$ for interaction, where each automaton has the set of secret states $Q_i^S$. Then the set of secret states of the system is $Q^S=Q\setminus Q^{NS}$, where $Q^{NS}=Q^{NS}_1\times\ldots\times Q^{NS}_n$. 
Let $\Delta\colon \ACT\trans\ACT\times\{\epsilon\}$ and $\Delta_R\colon \ACT\trans\{\epsilon\}\times\ACT$ and let $H_i=\langle \ACT_H,Q_H,\trans_H,Q^\circ_H\rangle$, where $\ACT_{H_i}=\Delta(\ACT_i)\cup\Delta_R(\ACT_i)$, be the two-way observer of $G_i$  for $1\leq i\leq n$.
If
\begin{eqnarray}
\nonumber &H_1\sync\ldots\sync H_n\trans[s]((X^1,X^1_R),\ldots,(X^n,X_R^n)):\\ 
          & [ (X^i\cap X_R^i)\not\subseteq Q_i^S \ \lor  (X^i\cap X^i_R)=\emptyset, \ \forall 1\leq i\leq n]
\end{eqnarray}
then $\SYSG$ is infinite-step opaque.}


\begin{theorem}\label{theom:synctwoInfinit}
\synctwoInfinit
\end{theorem}

\emph{Proof:} 

First, based on Proposition~\ref{propos:sync}, Proposition~\ref{pro:reverseNotisomorphic} and Lemma~\ref{lemm:renamingInsync} it holds that $H=\Delta(det(\SYSG))\sync\Delta_R(det(\SYSG_R)) \sqsubseteq	  [\Delta(det(G_1)\sync\ldots\sync det(G_n))\sync\Delta_R(det(G_{1,R})\sync\ldots\sync det(G_{n,R})) =  \Delta(det(G_1))\sync\ldots\sync \Delta(det(G_n))\sync\Delta_R(det(G_{1,R}))\sync\ldots\sync\Delta( det(G_{n,R}))=  \Delta(det(G_1))\sync \Delta_R(det(G_{1,R}))\sync\ldots\sync \Delta(det(G_n))\sync \Delta_R(det(G_{n,R})) = 
H_1\sync\ldots\sync H_n]$.

Assume that
\begin{eqnarray}\label{eq:them:infinitsync3}
 & H_1\sync\ldots\sync H_n\trans[s]((X^1,X^1_R),\ldots,(X^n,X_R^n))\ :\\
\nonumber & [(X^i\cap X^i_R)\not\subseteq Q_i^S \ \lor  (X^i\cap X_R^i)=\emptyset, \ \forall\ 1\leq i\leq n\ ]. 
\end{eqnarray}
 Since $\sync_\lor$ is used for interaction, from  $(X^i\cap X^i_R)\not\subseteq Q_i^S$ or $(X^i\cap X_R^i)=\emptyset$ for all $1\leq i\leq n$ it follows that $((X^1\cap X^1_R)\times\ldots\times(X^n\cap X^n_R))\not\subseteq Q^S$ or $((X^1\cap X^1_R)\times\ldots\times(X^n\cap X^n_R)) =\emptyset$. Therefore, equation~(\ref{eq:them:infinitsync3}) can be rewritten as 
\begin{eqnarray}\label{eq:them:infinitsync4}
\nonumber &H_1\sync\ldots\sync H_n\trans[s] ((X^1,X^1_R),\ldots,(X^n,X_R^n)) \ :\\
\nonumber &[((X^1\cap X^1_R)\times\ldots\times(X^n\cap X^n_R))\not\subseteq Q^S, \ \lor \\ & ((X^1\cap X^1_R)\times\ldots\times(X^n\cap X^n_R)) =\emptyset\ ]. 
\end{eqnarray}
 
Based on \Propn~\ref{pro:reverseNotisomorphic}  if  $H\trans[s](X,X_R)$ it holds that $H_1\sync \ldots\sync H_n\trans[s]((X^1,X^1_R),\ldots,(X^n,X_R^n))$ and $X_R\subseteq X^1_R\times\ldots \times X_R^n$  and based on  \Propn~\ref{propos:sync} it holds that $X=X^1\times\ldots\times X^n$.  This implies $(X\cap X_R)\subseteq (X^1\times\ldots\times X^n)\cap (X_R^1\times\ldots\times X_R^n)$.

Therefore, equation~(\ref{eq:them:infinitsync4}) can be rewritten as 
\begin{eqnarray*}
\nonumber&H\trans[s] (X,X_R) \ :\ [(X\cap X_R)\not\subseteq Q^S \ \lor\ (X\cap X_R)=\emptyset].
\end{eqnarray*}
Thus, the system is infinite-step opaque.\QEDA

\begin{figure}
	\centering
		\includegraphics{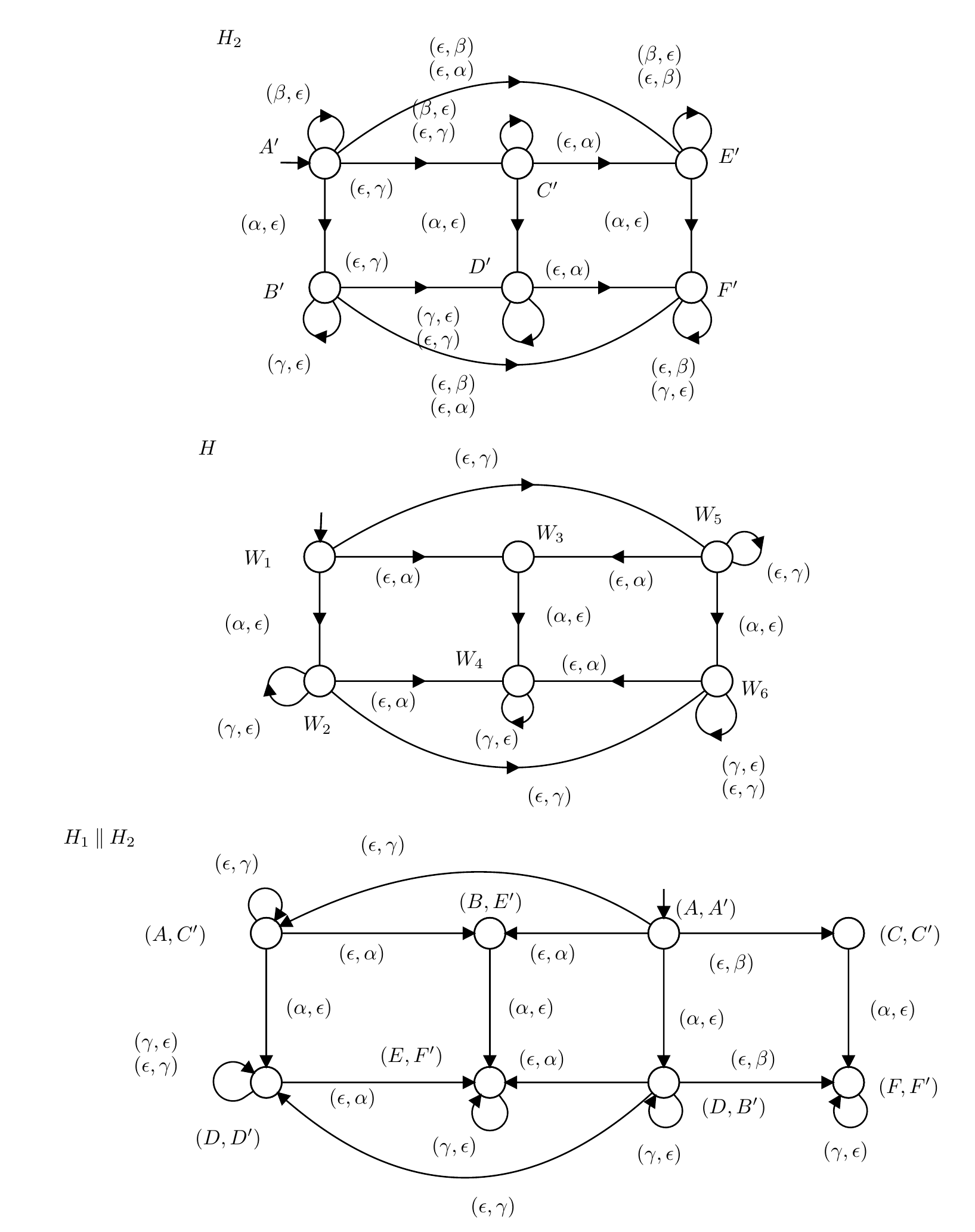}
\caption{Automata of Examples~\ref{ex:twowayInfiniLor} and~\ref{ex:twowayInfiniLand}.}\label{fig:twoSyncInf}
\end{figure}
\begin{example}\label{ex:twowayInfiniLor}
Consider the system $\SYSG=\{\tilde{G}_1,G_2\}$, where $\tilde{G}_1$ and $G_2$ are shown in Figure~\ref{fig:notion} and Figure~\ref{fig:syncIso} respectively. The set of secret states of $\tilde{G}_1$ is $\tilde{Q}^S_1=\{[s_1]\}$ and $G_2$ is $Q^S_2=\emptyset$. Assume $\sync_\lor$ is used for interaction. The two-way observer of $\tilde{G_1}$ is $\tilde{H}_1$ and the two-way observer of $G_2$ is $H_2$. Automaton $\tilde{H}_1$ is similar to $H_1$ shown in Figure~\ref{fig:notion}, where 

\begin{itemize}	
	\item $A=(\{s_0\},\{s_0,[s_1],[s_2],s_5\})$,
	\item $B=(\{s_0\},\{s_0\})$,
	\item $C=(\{s_0\},\{[s_1]\})$,		
	\item $D=(\{[s_1],[s_2]\},\{s_0,[s_1],[s_2],s_5\})$,
	\item $E=(\{[s_1],[s_2]\},\{s_0\})$,		
	\item $F=(\{[s_1],[s_2]\},\{[s_1]\})$,		
	\item $I=(\{s_5\},\{s_0,[s_1],[s_2],s_5\})$,		
	\item $J=(\{s_5\},\{s_0\})$,		
	\item $K=(\{s_5\},\{[s_1]\})$.
\end{itemize}

Figure~\ref{fig:twoSyncInf} shows $H_2$, where 

\begin{itemize}	
	\item $A'=(\{t_0\},\{t_0,t_1\})$,
	\item $B'=(\{t_1\},\{t_0,t_1\})$,
	\item $C'=(\{t_0\},\{t_1\})$,
	\item $D'=(\{t_1\},\{t_1\})$,
	\item $E'=(\{t_0\},\{t_0\})$,		
	\item $F'=(\{t_1\},\{t_0\})$.
\end{itemize}

The monolithic approach to verify infinite-step opacity of the system is to build $\SYSG=\tilde{G}_1\sync_\lor G_2$ and then build the two-way observer of the system $\SYSG$, which is shown in Figure~\ref{fig:twoSyncInf} as $H$. In the automaton $H$ the states are: 
\begin{itemize}	
	\item $W_1=((\{(s_0,t_0)\},\{(s_0,t_0),([s_1],t_1),([s_2],t_1)\})$,
	\item $W_2=(\{([s_1],t_1),([s_2],t_1)\},\{(s_0,t_0),([s_1],t_1)$ $,([s_2],t_1)\})$,
	\item $W_3=(\{(s_0,t_0)\},\{(s_0,t_0)\})$,		
	\item $W_4=(\{([s_1],t_1),([s_2],t_1)\},\{(s_0,t_0)\})$,
	\item $W_5=((\{(s_0,t_0)\},\{([s_1],t_1),([s_2],t_1)\})$,
	\item $W_6=(\{([s_1],t_1),([s_2],t_1)\},\{([s_1],t_1),([s_2],t_1)\})$,
\end{itemize}
 Let $W_i=(q_i,p_i)$ for automaton $H$. As it can be seen, $q_i\cap p_i=\{(s_0,t_0)\}\not\subseteq Q^S$ when $i=1,3$, and $q_i\cap p_i=\emptyset$ when $i=4,5$, and $q_i\cap p_i=\{([s_1],t_1),([s_2],t_1)\}\not\subseteq Q^S$ when $i=2,6$. Therefore, the system is infinite-step opaque. The modular two-way observer of the system is $H_1\sync H_2$ shown in Figure~\ref{fig:twoSyncInf}. The condition of Theorem~\ref{theom:synctwoInfinit} does not hold for the state $(F,F')$ in $H_1\sync H_2$ since for the state $F$ in $H_1$ we have $\{[s_1],[s_2]\}\cap\{[s_1]\}\subseteq Q^S_1$. Therefore, from $H_1\sync H_2$ it can not be concluded that the system is infinite-step opaque. 
\end{example}

\def\synctwoInfinitLand{Let $\mathcal{G}=G_1\sync_\land\ldots\sync_\land G_n$ be a nondeterministic system, with the set of secret states $Q_{\SYSG}^S=Q_1^S\times\cdots\times Q_n^S$. Let $\Delta\colon \ACT\trans\ACT\times\{\epsilon\}$ and $\Delta_R\colon \ACT\trans\{\epsilon\}\times\ACT$ and let $H_i=\langle \ACT_H,Q_H,\trans_H,Q^\circ_H\rangle$, where $\ACT_{H_i}=\Delta(\ACT_i)\cup\Delta_R(\ACT_i)$, be the two-way observer of $G_i$  for $1\leq i\leq n$.

If 
\begin{eqnarray}
\nonumber &H_1\sync\ldots\sync H_n\trans[s]((X^1,X^1_R),\ldots,(X^n,X_R^n)):\\ 
 & [\exists (X^i,X_R^i)\ \colon\ (X^i\cap X_R^i)\not\subseteq Q_i^S \ \lor  (X^i\cap X^i_R)=\emptyset\ ]
\end{eqnarray}
then $\SYSG$ is infinite-step opaque.}

\begin{corollary}\label{col:synctwoInfinit}
\synctwoInfinitLand
\end{corollary}

\emph{Proof:} 

First, based on Proposition~\ref{propos:sync}, Proposition~\ref{pro:reverseNotisomorphic} and Lemma~\ref{lemm:renamingInsync} it holds that $H=\Delta(det(\SYSG))\sync\Delta_R(det(\SYSG_R)) \sqsubseteq	  [\Delta(det(G_1)\sync\ldots\sync det(G_n))\sync\Delta_R(det(G_{1,R})\sync\ldots\sync det(G_{n,R})) =  \Delta(det(G_1))\sync\ldots\sync \Delta(det(G_n))\sync\Delta_R(det(G_{1,R}))\sync\ldots\sync\Delta( det(G_{n,R}))=  \Delta(det(G_1))\sync \Delta_R(det(G_{1,R}))\sync\ldots\sync \Delta(det(G_n))\sync \Delta_R(det(G_{n,R})) = 
H_1\sync\ldots\sync H_n]$.

Assume that
\begin{eqnarray}\label{eq:them:infinitsyncLand3}
\nonumber &H_1\sync\ldots\sync H_n\trans[s]((X^1,X^1_R),\ldots,(X^n,X_R^n))\ :\\
 & [\exists (X^i,X_R^i)\ \colon\ (X^i\cap X^i_R)\not\subseteq Q_i^S \ \lor  (X^i\cap X_R^i)=\emptyset ]\ . 
\end{eqnarray}
 Since $\sync_\land$ is used for interaction, from  there exist $(X^i,X_R^i)$ such that $(X^i\cap X^i_R)\not\subseteq Q_i^S$ or $(X^i\cap X_R^i)=\emptyset$ it follows that $((X^1\cap X^1_R)\times\ldots\times(X^n\cap X^n_R))\not\subseteq Q^S$ or $((X^1\cap X^1_R)\times\ldots\times(X^n\cap X^n_R)) =\emptyset$. Therefore, equation~(\ref{eq:them:infinitsyncLand3}) can be rewritten as 
\begin{eqnarray}\label{eq:col:infinitsyncLand4}
\nonumber &H_1\sync\ldots\sync H_n\trans[s]((X^1,X^1_R),\ldots,(X^n,X_R^n)) \ :\\
\nonumber &[((X^1\cap X^1_R)\times\ldots\times(X^n\cap X^n_R))\not\subseteq Q^S, \ \lor \\ & ((X^1\cap X^1_R)\times\ldots\times(X^n\cap X^n_R)) =\emptyset ]\ . 
\end{eqnarray}
 
Based on \Propn~\ref{pro:reverseNotisomorphic}  if  $H\trans[s](X,X_R)$ it holds that $H_1\sync \ldots\sync H_n\trans[s]((X^1,X^1_R),\ldots,(X^n,X_R^n))$ and $X_R\subseteq X^1_R\times\ldots \times X_R^n$  and based on  \Propn~\ref{propos:sync} it holds that $X=X^1\times\ldots\times X^n$.  This implies $(X\cap X_R)\subseteq (X^1\times\ldots\times X^n)\cap (X_R^1\times\ldots\times X_R^n)$. Therefore, equation~(\ref{eq:col:infinitsyncLand4}) can be rewritten as 
\begin{eqnarray*}
\nonumber&H\trans[s] (X,X_R) \ :\ [(X\cap X_R)\not\subseteq Q^S \ \lor\ (X\cap X_R)=\emptyset ] \ .
\end{eqnarray*}
Thus, the system is infinite-step opaque.\QEDA

\begin{example}\label{ex:twowayInfiniLand}
 Consider again the system $\SYSG=\{\tilde{G}_1,G_2\}$, where $\tilde{G}_1$ and $G_2$ are shown in Figure~\ref{fig:notion} and Figure~\ref{fig:syncIso} respectively. The set of secret states of $\tilde{G}_1$ is $\tilde{Q}^S_1=\{[s_1]\}$ and $G_2$ is $Q^S_2=\emptyset$. Now, assume $\sync_\land$ is used for interaction.  The problematic state $(F,F')$ in $H_1\sync H_2$ is no longer violating the condition of Corollary~\ref{col:synctwoInfinit} since for $F'$ we have $(\{t_0\}\cap\{t_1\})=\emptyset$. Therefore, the condition of Corollary~\ref{col:synctwoInfinit} holds for all the state of $H_1\sync H_2$. Thus, by analyzing the states of $H_1\sync H_2$, it can be concluded that the system is infinite-step opaque.
\end{example}

\subsection{Infinite-step opacity to nonblocking verification}\label{sec:infNon}

The main idea of this paper is to transform the opacity  verification problem to nonblocking verification and use the existing verification algorithms. To do so, the $\psi$-automata of the two-way observers of the system components need to be generated. In the following, Theorem~\ref{them:infiniteBlockingLor} and Corollary~\ref{cor:infiniteBlockingLand} show how $\psi$-automaton can be generated when $\sync_\lor$ and $\sync_\land$ are used for interaction, respectively. 

\def\infiniteBlockingLor{Let $\mathcal{G}=G_1\sync_\lor\ldots\sync_\lor G_n$ be a nondeterministic system, with the set of non-secret states $Q_{\SYSG}^{NS}=Q_1^{NS}\times\cdots\times Q_n^{NS}$ and the set of secret states $Q^S=Q\setminus Q^{NS}$. Let $\Delta\colon \ACT\trans\ACT\times\{\epsilon\}$ and $\Delta_R\colon \ACT\trans\{\epsilon\}\times\ACT$ and let $H_i=\langle \ACT_H,Q_H,\trans_H,Q^\circ_H\rangle$, where $\ACT_{H_i}=\Delta(\ACT_i)\cup\Delta_R(\ACT_i)$, be the two-way observer of $G_i$  for $1\leq i\leq n$.
If $H_{1,\psi_1}\sync \ldots \sync H_{n,\psi_n}$ is noblocking, where $X_{\psi,i}^i = \{(X^i,X^i_R)\in X_{H_i}\ |\ (X^i\cap X^i_R)\subseteq Q_i^S\}$, then the system $\SYSG$ is infinite-step opaque. }

\begin{theorem}\label{them:infiniteBlockingLor}
\infiniteBlockingLor
\end{theorem}

\emph{Proof:}  

Assume $H_{1,\psi_1}\sync \ldots \sync H_{n,\psi_n}$ is nonblocking. As all the states of $H_{i,\psi_i}$ for all $1\leq i\leq n$ except $\dumpstate$ are marked this means there does not exist $s\in (\Delta(\ACT)\cup \Delta_R(\ACT))^*$ such that $H_{1,\psi_1}\sync \ldots \sync H_{n,\psi_n}\trans[s]((X^1,X^1_R),\cdots,(X^n,X^n_R))\trans[\psi_i]\dumpstate$, which implies $(X^i,X^i_R)\not\in X_{\psi}^i$ for all $1\leq i\leq n$. Thus, for all $s\in (\Delta(\ACT)\cup \Delta_R(\ACT))^*$ it holds that 
$\Delta(det(G_1))\sync \Delta_R(det(G_{1,R}))\sync \ldots\sync \Delta(det(G_n))\sync \Delta_R(det(G_{n,R}))\trans[s]((X^1,X^1_R),\cdots,(X^n,X^n_R))$, where  $(X^i\cap X^i_R)\not\subseteq Q_i^S$ or $X^i\cap X^i_R=\emptyset$ for all $1\leq i\leq n$.  Therefore, based on Theorem~\ref{theom:synctwoInfinit} the system is infinite-step opaque.\QEDA


To generate $\psi$-automata of individual two-way observers, at the states of the two-way observers in which infinite-step opacity is violated the $\psi_i$-transition to blocking states are added. This transforms verification of infinite-step opacity to nonblocking problem. If the results of the verification is nonblocking then the system is infinite-step opaque. However, if the results is blocking the verification software tools like Supremica~\cite{AkeFabFloVah:03} will return a counter example. By the help of the counter example it is possible to verify if the blocking  is caused due to the over approximation. If this is the case, the problematic state can be removed and the system can be verified again, otherwise it can be concluded that the system is not infinite-step opaque. 

\begin{figure}
	\centering
		\includegraphics{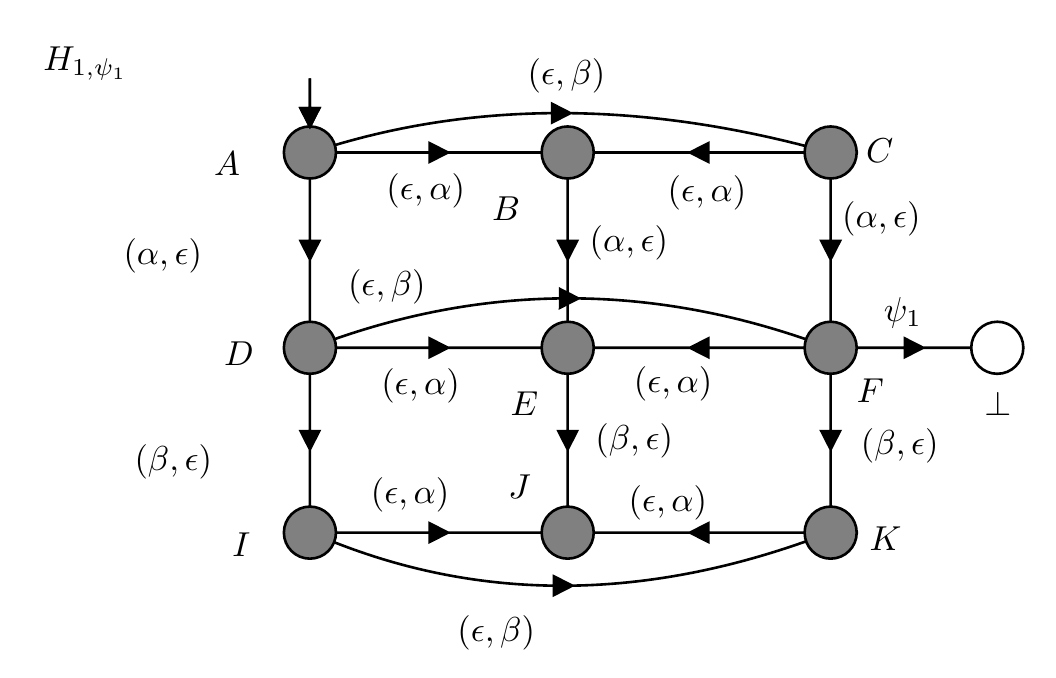}
\caption{Automata of Examples~\ref{ex:infNon} and~\ref{ex:1stepOpa}.}\label{fig:nonInfini}
\end{figure}

\begin{example}\label{ex:infNon}
Consider the system $\SYSG=\{G_1,G_2\}$, where $G_1$ and $G_2$ are shown in Figure~\ref{fig:notion} and Figure~\ref{fig:syncIso} respectively. The set of secret states of $G_1$ is $Q^S_1=\{s_1,s_3\}$ and $G_2$ is $Q^S_2=\emptyset$. Assume $\sync_\lor$ is used for interaction. At the first step of the compositional approach individual components are abstracted using opaque observation equivalence. This results in the abstracted system $\tilde{\SYSG}=\{\tilde{G}_1,G_2\}$ with $\tilde{Q}^S_1=\{[s_1]\}$. $\tilde{G}_1$ is shown in Figure~\ref{fig:notion}. After that, the two-way observers of the abstracted components are generated. As explained in Example~\ref{ex:twowayInfiniLor} the two-way observer of $\tilde{G_1}$ is $\tilde{H}_1$ and the two-way observer of $G_2$ is $H_2$ shown Figures~\ref{fig:notion} and~\ref{fig:twoSyncInf}.  Next, the infinite-step opacity verification is transformed to nonblocking verification by generating $\psi_i$-automaton for each $H_i$. For the two-way observer $H_1$ we have $X_{\psi_1}^1=\{F\}$ since for the state $F$ it holds that $\{[s_1],[s_2]\}\cap\{[s_1]\}\subseteq Q^S_1$. For the two-way observer $H_2$ we have $X_{\psi_2}^2=\emptyset$. Figure~\ref{fig:nonInfini} shows the $H_{1,\psi_1}$. The system $H_{1,\psi_1}\sync H_{2,\psi_2}$ is blocking since $H_{1,\psi_1}\sync H_{2,\psi_2}\trans[(\alpha,\epsilon)(\epsilon,\beta)](F,F')\trans[\psi_1]\dumpstate$. Thus, no conclusion can be drawn about the infinite-step opacity of the system. The problem of over approximation is caused by the reverse automata of the two-way observer. From $\Delta(det(G_1))\sync \Delta_R(det(G_{1,R}))\sync \Delta(det(G_2))\sync \Delta_R(det(G_{2,R}))\trans[(\alpha,\epsilon)(\epsilon,\beta)](F,F')=((\{[s_1],[s_2]\},\{[s_1]\}),(\{t_1\},\{t_0\}))$ it follows that $\Delta_R(G_{1,R})\sync \Delta_R(G_{2,R})\trans[(\epsilon,\beta)](\{[s_1]\},\{t_0\})$. Thus, to verify if this is a valid violation of opacity, it needs to be analyzed if state $([s_1],t_0)$ is a reachable state in $G_1\sync G_2$. Since $G_1\trans[\alpha][s_1]$ and $G_2\not\trans[\alpha]t_0$ it can be concluded that $([s_1],t_0)$ is not a reachable state in $G_1\sync G_2$. Thus, it can be concluded that the violation is due to the over approximation. Thus, the system is finite-step opaque.

\end{example}

\def\infiniteBlockingLand{Let $\mathcal{G}=G_1\sync_\land\ldots\sync_\land G_n$ be a nondeterministic system, with the set of secret states $Q_{\SYSG}^S=Q_1^S\times\cdots\times Q_n^S$. Let $\Delta\colon \ACT\trans\ACT\times\{\epsilon\}$ and $\Delta_R\colon \ACT\trans\{\epsilon\}\times\ACT$ and let $H_i=\langle \ACT_H,Q_H,\trans_H,Q^\circ_H\rangle$, where $\ACT_{H_i}=\Delta(\ACT_i)\cup\Delta_R(\ACT_i)$, be the two-way observer of $G_i$  for $1\leq i\leq n$.
If $H_{1,\psi}\sync \ldots \sync H_{n,\psi}$ is nonblocking, where $X_{\psi,i}^i = \{(X^i,X^i_R)\in X_{H_i}\ |\ (X^i\cap X^i_R)\subseteq Q_i^S\}$, then the system $\SYSG$ is infinite-step opaque. }

\begin{corollary}\label{cor:infiniteBlockingLand}
 \infiniteBlockingLand
\end{corollary}

\emph{Proof:}  

Assume $H_{1,\psi}\sync \ldots \sync H_{n,\psi}$ is nonblocking. As all the sates of $H_i$ for all $1\leq i\leq n$ are marked this means there does not exist $s\in (\Delta(\ACT)\cup \Delta_R(\ACT))^*$ such that $H_{1,\psi}\sync \ldots \sync H_{n,\psi}\trans[s]((X_1,Y_1),\cdots,(X_n,Y_n))\trans[\psi]\dumpstate$, which implies there exists $(X_i,Y_i)$ such that $(X_i,Y_i)\not\in X_{\psi}^i$. Thus, for all $s\in (\Delta(\ACT)\cup \Delta_R(\ACT))^*$ it holds that 
$\Delta(det(G_1))\sync \Delta_R(det(G_{1,R}))\sync \ldots\sync \Delta(det(G_n))\sync \Delta_R(det(G_{n,R}))\trans[s]((X^1,X_R^1),\cdots,(X^n,X^n_R))$, where  $(X^i\cap X^i_R)\not\subseteq Q_i^S$ or $(X^i\cap X^i_R)=\emptyset$ for some $1\leq i\leq n$. Based on Corollary~\ref{col:synctwoInfinit} it holds that the system is infinite-step opaque.\QEDA


\section{Compositional $K$-step opacity verification}\label{sec:Kstep} 

This section discusses the compositional approach for $K$-step opacity verification. 
Since infinite-step opacity is a special case of $K$-step opacity, in this section we mainly focus on $K$-step opacity and we discuss throughout how the results can be extended to handle infinite-step opacity verification. 
As in the case of current-state opacity, the first step of compositional  $K$-step opacity verification is to abstract the individual components. This is described in \Sect~\ref{sec:Kobs}. 
Next, \Sect~\ref{sec:syncK} presents the construction of two-way observers of individual components and \Sect~\ref{sec:nonK} shows how  $K$-step opacity verification can be transformed to nonblocking verification.

\subsection{Opaque observation equivalence}\label{sec:Kobs}

The idea of $K$-step opacity is that the intruder cannot determine if the secret state was reached within $K$-step prior to the current state. As opaque observation equivalence only merges states with the same secrecy property and the same future behavior, $K$-step opacity is preserved by opaque observation equivalence, step (i) in \Fig~\ref{fig:composAlg}. This is shown in \Thm~\ref{them:KstepOE}. 

\def\KstepOE{
  Let $\mathcal{G}=\{G_1,\ldots, G_n\}$ be a non-deterministic system with $\sync_\lor$ for interaction, where each automaton has  set of secret states $Q_i^S$. Hence, the set of secret states of the system is $Q^S=Q\setminus Q^{NS}$, where $Q^{NS}=Q^{NS}_1\times\ldots\times Q^{NS}_n$.  Let $\sim$ be an opaque observation equivalence on $G_1$ such that $\tilde{\SYSG}=\{\tilde{G_1},G_2,\ldots,G_n\}$. Then $\SYSG$ is $K$-step opaque if and only if $\tilde{\SYSG}$ is $K$-step opaque.}
	
\begin{theorem}\label{them:KstepOE}
\KstepOE
\end{theorem}
\emph{Proof:}  

Consider  $G_2\sync\ldots\sync G_n=T=\langle \ACT_T,Q_T,\trans,Q^\circ_T\rangle$ with the set of secret states $Q^S_T=Q^S_2\times \ldots\times Q^T_n$ and non secret states $Q_T^{NS}=Q_T\setminus Q_T^S$. It suffices to show that if $G_1\sync_\lor T$ is not $K$-step opaque then $\tilde{G}_1\sync_\lor T$ is not $K$-step opaque either and vice versa. 
\begin{enumerate}
\item First assume that $G_1\sync_\lor T$ is not $K$-step opaque. Then there exists $(x^0_{G_1},x^0_T)\in Q^\circ_{G_1}\times Q^\circ_T$ and $st\in\LANG(G_1\sync_\lor T,(x^0_{G_1},x^0_T))$ such that $(x^0_{G_1},x^0_T)\ttrans[s](x_1,x_T)$, and $x_1\in Q_{1}^S$ or $x_T\in  Q_T^S$ and $|t|\leq K$, and  one of the following holds.
\begin{itemize}
\item For all $(x'_1,x'_T)\in Q^{NS}_1\times Q^{NS}_T$ such that $G_1\sync_\lor T\ttrans[s](x'_1,x'_T)$ then $G_1\sync_\lor T\ttrans[s](x'_1,x'_T)\not\ttrans[t]$. 
\item There does not exist $(x'_1,x'_T)\in Q^{NS}_1\times Q^{NS}_T$ such that $G_1\sync_\lor T\ttrans[s](x'_1,x'_T)$.

\end{itemize} 
Since $G_1\sim \tilde{G}_1$ it holds that $\tilde{G}_1\sync_\lor T \sim G_1\sync_\lor T$. This means, if $G_1\sync_\lor T\ttrans[s](x'_1,x'_T)\not\ttrans[t]$ then $\tilde{G}_1\sync_\lor T\ttrans[s]([x'_1],x'_T)\not\ttrans[t]$, where  $x'_1\in[x_1']$ and $(x'_1,x'_T)\in Q^{NS}_1\times Q^{NS}_T$ if and only if $([x'_1],x'_T)\in \tilde{Q}^{NS}_1\times Q^{NS}_T$. Thus, $\tilde{G}_1\sync_\lor T$ is not $K$-step opaque either. 

 Now consider the case where $G_1\sync_\lor T$ is not $K$-step opaque because $(x^0_G,x^0_T)\ttrans[s](x_1,x_T)$, and $x_1\in Q_1^S$ or $x_T\in  Q_T^S$ and there does not exist $(x'_1,x'_T)\in Q^{NS}_1\times Q^{NS}_T$ such that $G_1\sync_\lor T\ttrans[s](x'_1,x'_T)$.
The proof for this part is the same as the proof for \Thm~\ref{thm:OpaqueOECS} in \Sect~\ref{sec:obsEqCurrent}, proof of opaque observation equivalence preserves current-state opacity, and consequently omitted.

\item If $\tilde{G}_1\sync_\lor T$ is not $K$-step opaque. Then there exists $([x_{G_1}^0],x_T^0)\in \tilde{Q}^\circ_{G_1}\times Q_T^\circ$ and $st\in\LANG(\tilde{G}_1\sync_\lor T,([x^0_{G_1}],x^0_T))$ such that $([x^0_{G_1}],x^0_T)\ttrans[s]([x_1],x_T)$, and $[x_1]\in \tilde{Q}_1^S$ or $x_T\in  Q_T^S$ and $|t|\leq K$, and one of the following cases holds.
\begin{itemize}
\item For all $([x'_1],x'_T)\in \tilde{Q}^{NS}_1\times Q^{NS}_T$ such that $\tilde{G}_1\sync_\lor T\ttrans[s]([x'_1],x'_T)$ then $\tilde{G}_1\sync_\lor T\ttrans[s]([x'_1],x'_T)\not\ttrans[t]$.
\item There does not exist $([x'_1],x'_T)\in \tilde{Q}^{NS}_{1}\times Q^{NS}_T$ such that $\tilde{G}_1\sync_\lor T\ttrans[s]([x'_1],x'_T)$. 
\end{itemize} 
Since $G_1\sim \tilde{G}_1$ it holds that $\tilde{G}_1\sync_\lor T \sim G_1\sync_\lor T$. This means, if $\tilde{G}_1\sync_\lor T\ttrans[s]([x'_1],x'_T)\not\ttrans[t]$ then $G_1\sync_\lor T\ttrans[s](x'_1,x'_T)\not\ttrans[t]$, where  $x'_1\in[x_1']$ and $(x'_1,x'_T)\in Q^{NS}_1\times Q^{NS}_T$ if and only if $([x'_1],x'_T)\in \tilde{Q}^{NS}_1\times Q^{NS}_T$. Thus, $G_1\sync_\lor T$ is not $K$-step opaque either.   

Now consider the case when $\tilde{G}_1\sync_\lor T$ is not $K$-step opaque because $([x^0_{G_1}],x^0_T)\ttrans[s]([x_1],x_T)$, and $[x_1]\in \tilde{Q}_1^S$ or $x_T\in  Q_T^S$ and there does not exist $([x'_1],x'_T)\in \tilde{Q}^{NS}_1\times Q^{NS}_T$ such that $\tilde{G}_1\sync_\lor T\ttrans[s]([x'_1],x'_T)$.
The proof for this part is the same as the proof for \Thm~\ref{thm:OpaqueOECS} in \Sect~\ref{sec:obsEqCurrent}, proof of opaque observation equivalence preserves current-state opacity, and consequently omitted. \QEDA
\end{enumerate}

\Thm~\ref{them:KstepOE} shows that if $x\ttrans[s]$ in $G$, then $[x]\ttrans[s]$ in $\tilde{G}$. As  opaque observation equivalence only merges states with the same future behavior, the lengths of the strings remain consistent after abstraction. Therefore, $K$-step opacity is preserved by opaque observation equivalence. 

\def\KstepOELand{
Let $\mathcal{G}=\{G_1,\ldots, G_n\}$ be a non-deterministic system with $\sync_\land$ for interaction and with the set of secret states  $Q^S=Q_1^S\times \ldots\times Q_n^S$.  Let $\sim$ be an opaque observation equivalence on $G_1$ such that $\tilde{\SYSG}=\{\tilde{G_1},G_2,\ldots,G_n\}$. Then $\SYSG$ is $K$-step opaque if and only if $\tilde{\SYSG}$ is $K$-step  opaque.}
	
\begin{corollary}\label{cor:KstepOE}
\KstepOELand
\end{corollary}
\emph{Proof:}

Consider $G_2\sync\ldots\sync G_n=T=\langle \ACT_T,Q_T,\trans,Q^\circ_T\rangle$. It suffices to show that if $G_1\sync_\land T$ is not $K$-step opaque then $\tilde{G}_1\sync_\land T$ is not $K$-step opaque either and vice versa. 
\begin{enumerate}
\item  First assume that $G_1\sync_\land T$ is not $K$-step opaque. Then there exists $(x^0_{G_1},x^0_T)\in Q^\circ_{G_1}\times Q^\circ_T$ and $st\in\LANG(G_1\sync_\land T,(x^0_{G_1},x^0_T))$ such that $(x^0_{G_1},x^0_T)\ttrans[s](x_1,x_T)$, $(x_1,x_T)\in Q_1^S\times   Q_T^S$ and $|t|\leq K$, and  one of the following holds.
\begin{itemize}
\item For all $(x'_1,x'_T)$ such that $x'_1\in Q^{NS}_1$ or $x'_T\in Q^{NS}_T$ and $G_1\sync_\land T\ttrans[s](x'_1,x'_T)$ then $G_1\sync_\land T\ttrans[s](x'_1,x'_T)\not\ttrans[t]$. 
\item There does not exist $x'_1\in Q^{NS}_1$ or $x'_T\in Q^{NS}_T$ such that $G_1\sync_\land T\ttrans[s](x'_1,x'_T)$.

\end{itemize} 
Since $G_1\sim \tilde{G}_1$ it holds that $\tilde{G}_1\sync_\land T \sim G_1\sync_\land T$. This means, if $G_1\sync_\land T\ttrans[s](x'_1,x'_T)\not\ttrans[t]$ then $\tilde{G}_1\sync_\land T\ttrans[s]([x'_1],x'_T)\not\ttrans[t]$, where  $x'_1\in[x_1']$ and $x'_1\in Q^{NS}_1$ if and only if $[x'_1]\in \tilde{Q}^{NS}_1$. Thus, $\tilde{G}_1\sync_\land T$ is not $K$-step opaque either.

 Now consider the case, where $(x_1,x_T)\in Q_{G_1}\times Q_T$ such that $G_1\sync_\land T\ttrans[s](x_1,x_T)$ and $(x_1,x_T)\in Q_1^S\times  Q_T^S$, and there does not exist $x'_1\in Q^{NS}_1$ or $x'_T\in Q^{NS}_T$ such that $G_1\sync_\land T\ttrans[s](x'_1,x'_T)$. The proof for this part is the same as the proof for Corollary~\ref{cor:obsEqCSLand}, proof of opaque observation equivalence preserves current-state opacity, and consequently omitted. 

 Therefore, it can be concluded that $\tilde{G}_1\sync_\land T$ is not $K$-step opaque.

\item If $\tilde{G}_1\sync_\land T$ is not $K$-step opaque. Then there exists $([x_{G_1}^0],x_T^0)\in \tilde{Q}^\circ_{G_1}\times Q_T^\circ$ and $st\in\LANG(\tilde{G}_1\sync_\land T,([x^0_G],x^0_T))$ such that $([x^0_G],x^0_T)\ttrans[s]([x_1],x_T)$,  $([x_1],x_T)\in \tilde{Q}_{G_1}^S\times  Q_T^S$ and $|t|\leq K$, and one of the following cases holds.
\begin{itemize}
\item For all $([x'_1],x'_T)\in \tilde{Q}_{G_1}\times Q_T$  such that $[x'_1]\in \tilde{Q}^{NS}$ or $x'_T\in  Q^{NS}_T$ and $\tilde{G}_1\sync_\land T\ttrans[s]([x'_1],x'_T)$ then $\tilde{G}_1\sync_\land T\ttrans[s]([x'_1],x'_T)\not\ttrans[t]$.
\item There does not exist $([x'_1],x'_T)\in \tilde{Q}^{NS}_1\times Q^{NS}_T$ such that $\tilde{G}_1\sync_\land T\ttrans[s]([x'_1],x'_T)$. 
\end{itemize} 
Since $G_1\sim \tilde{G}_1$ it holds that $\tilde{G}_1\sync_\land T \sim G_1\sync_\lor T$. This means, if $\tilde{G}_1\sync_\land T\ttrans[s]([x'_1],x'_T)\not\ttrans[t]$ then $G_1\sync_\land T\ttrans[s](x'_1,x'_T)\not\ttrans[t]$, where  $x'_1\in[x_1']$ and $x'_1\in Q^{NS}_1$ if and only if $[x'_1]\in \tilde{Q}^{NS}_1$. Thus, $G\sync_\land T$ is not $K$-step opaque either.

Again the proof for this part is the same as the proof for Corollary~\ref{cor:obsEqCSLand}, proof of opaque observation equivalence preserves current-state opacity, and consequently omitted. 

 Therefore, it can be concluded that the $G\sync_\land T$ is not $K$-step opaque.\QEDA
\end{enumerate}

\subsection{Synchronization of two-way observers}\label{sec:syncK}

 To verify  $K$-step opacity similar to infinite-step opacity the two-way observers of the system components need to be generated. As explained in Section~\ref{sec:syncInf} synchronization of the two-way observers of individual components provides an over approximation of the state space of the monolithic two-way observer of the  system. Thus, if the verification of the over approximated two-way observer results in $K$-step opacity it can be concluded that the original system is also $K$-step opaque. However,  if the result shows violation of the $K$-step opacity it needs to be investigated to confirm that the violation is not the result of reaching an unreachable state due to over approximation. 

\def\syncTwoK{Let $\mathcal{G}=G_1\sync_\lor\ldots\sync_\lor G_n$ be a  non-deterministic system, where each automaton has  set of  secret states $Q_i^S$. Hence, the set of secret states of the system is $Q^S= Q\setminus Q^{NS}$, where $Q^{NS}=Q_1^{NS}\times\ldots\times Q_n^{NS}$. Let $\Delta\colon \ACT\trans\ACT\times\{\epsilon\}$ and $\Delta_R\colon \ACT\trans\{\epsilon\}\times\ACT$ and let $H_i=\langle \ACT_H,Q_H,\trans_H,Q^\circ_H\rangle$, where $\ACT_{H_i}=\Delta(\ACT_i)\cup\Delta_R(\ACT_i)$, be the two-way observer of $G_i$ and let $H=H_1\sync\ldots\sync H_n$. Let $P_{\Delta}:\ACT_{H}\trans \Delta_R(\ACT)$. If for any string $s\in\LANG(H_1\sync\ldots\sync H_n)$ such that $H_1\sync\ldots\sync H_n\trans[s]((X^1,X_R^1),\ldots,(X^n,X_R^n))$ we have 
 $$
\begin{array}{@{}r@{\quad}l@{}}
  &[\exists(X^i\cap X_R^i)\ :\ (X^i\cap X_R^i)\subseteq Q_i^S \land\ (X^i\cap X_R^i)\not=\emptyset]\\ & \Rightarrow  |P_\Delta(s)|> K,
 \end{array}
$$
then the system $\SYSG$ is $K$-step opaque.}
\begin{theorem}\label{them:syncTwoK}
\syncTwoK
\end{theorem}
\emph{Proof:} 

First, based on Proposition~\ref{propos:sync}, Proposition~\ref{pro:reverseNotisomorphic} and Lemma~\ref{lemm:renamingInsync} it holds that $\Delta(det(\SYSG))\sync\Delta_R(det(\SYSG_R))\sqsubseteq	 [\Delta(det(G_1)\sync\ldots\sync det(G_n))\sync\Delta_R(det(G_{1,R})\sync\ldots\sync det(G_{n,R})) =  \Delta(det(G_1))\sync\ldots\sync \Delta(det(G_n))\sync\Delta_R(det(G_{1,R}))\sync\ldots\sync\Delta( det(G_{n,R}))=  \Delta(det(G_1))\sync \Delta_R(det(G_{1,R}))\sync\ldots\sync \Delta(det(G_n))\sync \Delta_R(det(G_{n,R})) = 
H_1\sync\ldots\sync H_n]$.

Assume that for any string $s\in\LANG(H_1\sync\ldots\sync H_n)$ such that $H_1\sync\ldots\sync H_n\trans[s]((X_1,Y_1),\ldots,(X_n,Y_n))$  we have

\begin{eqnarray}\label{app:eq:Kmodular1}
 \nonumber &[\exists(X^i\cap X_R^i)\ :\ (X^i\cap X_R^i)\subseteq Q_i^S \land\ (X^i\cap X_R^i)\not=\emptyset]\\ &\Rightarrow  |P_\Delta(s)|> K,
 \end{eqnarray}

 Since $\sync_\lor$ is used for interaction, from $[\exists(X^i\cap X_R^i)\ :\ (X^i\cap X_R^i)\subseteq Q_i^S \land\ (X^i\cap X_R^i)\not=\emptyset]$ it holds that $[((X^1\cap X^1_R)\times\ldots\times(X^n\cap X^n_R))\subseteq Q^S\land ((X^1\cap X^1_R)\times\ldots\times(X^n\cap X^n_R))\not=\emptyset]$. Since $((X^1\cap X^1_R)\times\ldots\times(X^n\cap X^n_R))=((X^1\times\ldots\times X^n)\cap(X^1_R\times \ldots\times X^n_R))$ equation~\ref{app:eq:Kmodular1} can be rewritten as  

$$
\begin{array}{@{}r@{\quad}l@{}}
  &[((X^1\times\ldots\times X^n)\cap(X_R^1\times\ldots\times X^n_R))\subseteq Q^S \\&\land\ ((X^1\times\ldots\times X^n)\cap(X_R^1\times\ldots\times X^n_R))\not =\emptyset] \Rightarrow  |P_\Delta(s)|> K,
 \end{array}
$$

Based on \Propn~\ref{pro:reverseNotisomorphic}  if  $H\trans[s](X,X_R)$ it holds that $H_1\sync \ldots\sync H_n\trans[s]((X^1,X^1_R),\ldots,(X^n,X_R^n))$ and $X_R\subseteq X^1_R\times\ldots \times X_R^n$  and based on  \Propn~\ref{propos:sync} it holds that $X=X^1\times\ldots\times X^n$.  This implies $(X\cap X_R)\subseteq (X^1\times\ldots\times X^n)\cap (X_R^1\times\ldots\times X_R^n)$. Thus, there are two possibilities for  $(X\cap X_R)$:
\begin{itemize}
\item If $(X\cap X_R)=\emptyset$ then it can be concluded that the system is $K$-step opaque.
\item If $(X\cap X_R)\not=\emptyset$ then as $((X^1\times\ldots\times X^n)\cap (X_R^1\times\ldots\times X_R^n))\subseteq Q^S$ and $(X\cap X_R)\subseteq ((X^1\times \ldots\times X^n)\cap (X_R^1\times \ldots\times X_R^n))$ it follows that $(X\cap X_R)\subseteq Q^S$. Thus, we have

\begin{eqnarray}\label{eq:them:ksync5}
\nonumber [ (X\cap X_R)\subseteq Q^S \land (X\cap X_R)\not=\emptyset ]\  \Rightarrow\ |P_\Delta(s)|> K, \
\end{eqnarray}
which means the system is $K$-step opaque.\QEDA
\end{itemize}

\Thm~\ref{them:syncTwoK} establishes a sufficient condition for $K$-step opacity verification when $\sync_\lor$ is used for synchronization. Essentially, the theorem looks at all   reachable states of the modular two-way observer, which contain states in which the (non-empty) intersection of the two components lies entirely in the set of secret states. If such states are reached after up to $K$ observations, then the system is $K$-step opaque.


\begin{example}\label{ex:twowayInfiniLor2}
Consider the system $\SYSG=\{\tilde{G}_1,G_2\}$, where $\tilde{G}_1$ and $G_2$ are shown in \Fig~\ref{fig:notion} and \Fig~\ref{fig:syncIso}, respectively. The set of secret states of $\tilde{G}_1$ is $\tilde{Q}^S_1=\{[s_1]\}$ and that of $G_2$ is $Q^S_2=\emptyset$. The modular two-way observer of the system is $\{H_1, H_2\}$ and the monolithic two-way observer of the system is $H$, shown  in \Fig~\ref{fig:twoSyncInf}.  Assume $\sync_\lor$ is used for interaction and 1-step opacity needs to be verified. Let $W_i=(q_i,p_i)$ for automaton $H$. As it can be seen, $q_i\cap p_i=\{(s_0,t_0)\}\not\subseteq Q^S$ when $i=1,3$, and $q_i\cap p_i=\emptyset$ when $i=4,5$, and $q_i\cap p_i=\{([s_1],t_1),([s_2],t_1)\}\not\subseteq Q^S$ when $i=2,6$. Therefore, the system is 1-step opaque and in fact infinite-step opaque. However, the condition of \Thm~\ref{them:syncTwoK} does not hold for the state $(F,F')$ since $H_1\sync H_2\trans[(\alpha,\epsilon)(\epsilon,\beta)](F,F')$ and for state $F$ in $H_1$ we have that $\{[s_1],[s_2]\}\cap\{[s_1]\}\subseteq Q^S_1$ but $|P_\Delta((\alpha,\epsilon)(\epsilon,\beta))|=1\not>1$. Therefore, from $H_1\sync H_2$, it cannot be concluded that the system is 1-step opaque.  
\end{example}

Corollary~\ref{col:syncTwoKland} provides a similar result as \Thm~\ref{them:syncTwoK} when interaction is done using $\sync_\land$.
\def\syncTwoKLand{Let $\mathcal{G}=G_1\sync_\land\ldots\sync_\land G_n$ be a   non-deterministic system, where each automaton has  set of  secret states $Q_i^S$. Hence, the set of secret states of the system is $Q^S= Q_1^{S}\times\ldots\times Q_n^{S}$. Let $\Delta\colon \ACT\trans\ACT\times\{\epsilon\}$ and $\Delta_R\colon \ACT\trans\{\epsilon\}\times\ACT$ and let $H_i=\langle \ACT_H,Q_H,\trans_H,Q^\circ_H\rangle$, where $\ACT_{H_i}=\Delta(\ACT_i)\cup\Delta_R(\ACT_i)$, be the two-way observer of $G_i$ and let $H=H_1\sync\ldots\sync H_n$. Let $P_{\Delta}:\ACT_{H}\trans \Delta_R(\ACT)$. If for any string $s\in\LANG(H_1\sync\ldots\sync H_n)$ such that $H_1\sync\ldots\sync H_n\trans[s]((X_1,Y_1),\ldots,(X_n,Y_n))$ we have
 $$
\begin{array}{@{}r@{\quad}l@{}}
  &[ (X^i\cap X_R^i)\subseteq Q_i^S \land\ (X^i\cap X_R^i)\not=\emptyset, \forall 1\leq i\leq n]\\ & \Rightarrow  |P_\Delta(s)|> K,
 \end{array}
$$
then the system $\SYSG$ is $K$-step opaque.}

\begin{corollary}\label{col:syncTwoKland}
\syncTwoKLand
\end{corollary}
\emph{Proof:}

First, based on \Propn~\ref{propos:sync}, \Propn~\ref{pro:reverseNotisomorphic} and Lemma~\ref{lemm:renamingInsync} it holds that $\Delta(det(\SYSG))\sync\Delta_R(det(\SYSG_R))\sqsubseteq	 [\Delta(det(G_1)\sync\ldots\sync det(G_n))\sync\Delta_R(det(G_{1,R})\sync\ldots\sync det(G_{n,R})) =  \Delta(det(G_1))\sync\ldots\sync \Delta(det(G_n))\sync\Delta_R(det(G_{1,R}))\sync\ldots\sync\Delta( det(G_{n,R}))=  \Delta(det(G_1))\sync \Delta_R(det(G_{1,R}))\sync\ldots\sync \Delta(det(G_n))\sync \Delta_R(det(G_{n,R})) = 
H_1\sync\ldots\sync H_n]$.
Assume that for any string $s\in\LANG(H_1\sync\ldots\sync H_n)$ such that $H_1\sync\ldots\sync H_n\trans[s]((X_1,Y_1),\ldots,(X_n,Y_n))$ 

\begin{eqnarray}\label{app:eq:kmodularsync2}
 \nonumber &[(X^i\cap X_R^i)\subseteq Q_i^S \land\ (X^i\cap X_R^i)\not=\emptyset, \forall 1\leq i\leq n]\\ &\Rightarrow  |P_\Delta(s)|> K,
 \end{eqnarray}

 Since $\sync_\land$ is used for interaction, from  $(X^i\cap X^i_R)\subseteq Q_i^S$ and $(X^i\cap X_R^i)\not=\emptyset$ for all $1\leq i\leq n$ it holds that $(X^1\cap X^1_R)\times\ldots\times(X^n\cap X^n_R)\subseteq Q^S$ and $(X^1\cap X^1_R)\times\ldots\times(X^n\cap X^n_R)\not=\emptyset$. Since $(X^1\cap X^1_R)\times\ldots\times(X^n\cap X^n_R)=(X^1\times\ldots\times X^n)\cap(X^1_R\times\ldots\times X^n_R)$ equation~\ref{app:eq:kmodularsync2} can be rewritten as $[(X^1\times\ldots\times X^n)\cap(X_R^1\times\ldots\times X^n_R)\subseteq Q^S \land\ (X^1\times\ldots\times X^n)\cap(X_R^1\times\ldots\times X^n_R)\not =\emptyset] \Rightarrow  |P_\Delta(s)|> K$.


Based on \Propn~\ref{pro:reverseNotisomorphic}  if  $H\trans[s](X,X_R)$ it holds that $H_1\sync \ldots\sync H_n\trans[s]((X^1,X^1_R),\ldots,(X^n,X_R^n))$ and $X_R\subseteq X^1_R\times\ldots \times X_R^n$  and based on  \Propn~\ref{propos:sync} it holds that $X=X^1\times\ldots\times X^n$.  This implies $(X\cap X_R)\subseteq (X^1\times\ldots\times X^n)\cap (X_R^1\times\ldots\times X_R^n)$. 
Thus, there are two possibilities for  $(X\cap X_R)$:
\begin{itemize}
\item If $(X\cap X_R)=\emptyset$ then it can be concluded that the system is $K$-step opaque.
\item If $(X\cap X_R)\not=\emptyset$ then from $(X^1\times\ldots\times X^n)\cap (X_R^1\times\ldots\times X_R^n)\subseteq Q^S$ and $(X\cap X_R)\subseteq (X^1\times\ldots\times X^n)\cap (X_R^1\times\ldots\times X_R^n)$ it follows that $(X\cap X_R)\subseteq Q^S$. Thus, we have

\begin{eqnarray}
\nonumber [ (X\cap X_R)\subseteq Q^S \land (X\cap X_R)\not=\emptyset ]\  \Rightarrow\ |P_\Delta(s)|> K, \
\end{eqnarray}
which means the system is $K$-step opaque.\QEDA
\end{itemize}

\begin{example}\label{ex:twowayInfiniLor3}
Consider the system $\SYSG=\{\tilde{G}_1,G_2\}$, where $\tilde{G}_1$ and $G_2$ are shown in \Fig~\ref{fig:notion} and \Fig~\ref{fig:syncIso}, respectively. The set of secret states of $\tilde{G}_1$ is $\tilde{Q}^S_1=\{[s_1]\}$ and that of $G_2$ is $Q^S_2=\emptyset$.  The modular two-way observer of the system is $\{H_1, H_2\}$ and the monolithic two-way observer of the system is $H$, shown  in \Fig~\ref{fig:twoSyncInf}. Assume $\sync_\land$ is used for interaction and 1-step opacity needs to be verified. \footnote{Since $Q^S_2=\emptyset$ and $\sync_\land$ is used for interaction the system is trivially infinite-step opaque and also 1-step opaque. However, it is used to explain the  condition of Corollary~\ref{col:syncTwoKland}.}  The condition of Corollary~\ref{col:syncTwoKland}  holds for all states of $H_1\sync H_2$ including state $(F,F')$.  Even though for state $F$ in $H_1$ we have that $\{[s_1],[s_2]\}\cap\{[s_1]\}\subseteq Q^S_1$ but since for $F'$ we have ${t_1}\cap{t_0}=\emptyset$, state $(F,F')$ does not violate the condition of Corollary~\ref{col:syncTwoKland}. Therefore, in contrast to \Examp~\ref{ex:twowayInfiniLor2} from $H_1\sync H_2$, it can be concluded that the system is 1-step opaque.  
\end{example}

\subsection{$K$-step opacity to nonblocking verification}\label{sec:nonK}

The main idea of this paper is to transform opacity verification to nonblocking verification and use existing nonblocking verification algorithms. 
To transform $K$-step opacity verification to nonblocking verification, the $\psi$-automata of the two-way observers need to be built. First, the states of the  two-way observers that violate  $K$-step opacity are identified and from them  transitions to  blocking states are added, step (iii) and (iv) in \Fig~\ref{fig:composAlg}. 

\def\OpBlockingKlor{Let $\mathcal{G}=G_1\sync_\lor\ldots\sync_\lor G_n$ be a non-deterministic system, with  set of non-secret states $Q_{\SYSG}^{NS}=Q_1^{NS}\times\cdots\times Q_n^{NS}$. Let $\Delta\colon \ACT\trans\ACT\times\{\epsilon\}$ and $\Delta_R\colon \ACT\trans\{\epsilon\}\times\ACT$ and let $H_i=\langle \ACT_H,Q_H,\trans_H,Q^\circ_H\rangle$, where $\ACT_{H_i}=\Delta(\ACT_i)\cup\Delta_R(\ACT_i)$, be the two-way observer of $G_i$  for $1\leq i\leq n$. Let $P_{i,\Delta}:\Sigma_{H_i}\to \Delta_R(\ACT_i)$.
If $H_{1,\psi_1}\sync \ldots \sync H_{n,\psi_n}$ is nonblocking then  $\SYSG$ is $k$-step opaque, where $X_{\psi_i}^i = \{\ H_i\trans[s](X^i, X_R^i)\ |\ (X^i\cap X_R^i)\subseteq Q_i^S \land\ (X^i\cap X_R^i)\not =\emptyset\ \land |P_{i,\Delta}(s)|\leq K\}$. }
\begin{theorem}\label{them:OpBlockingKlor}
\OpBlockingKlor
\end{theorem}
\emph{Proof:}

If $H_{1,\psi_1}\sync \ldots \sync H_{n,\psi_n}$ is nonblocking it means that for all $s\in (\Delta(\ACT)\cup \Delta_R(\ACT))^*$ such that $H_{1,\psi_1}\sync \ldots \sync H_{n,\psi_n}\trans[s]((X^1,X_R^1),\ldots,(X^n,X_R^n))$  there does not exist $\psi_i$ such that $((X^1,X_R^1),\ldots,(X_n,X_R^n))\not\trans[\psi_i]\dumpstate$. This means $(X^i,X_R^i)\not\in X_{\psi_i}^i$ for all $1\leq i\leq n$. If $(X^i,X_R^i)\not\in X_{\psi_i}^i$, there are two possibilities:
\begin{itemize}
\item   $(X^i\cap X_R^i)\not\subseteq Q^S_i$ for all $1\leq i\leq n$. This means for all $s\in (\Delta(\ACT)\cup \Delta_R(\ACT))^*$ such that $H_{1}\sync \ldots \sync H_{n}\trans[s]((X^1,X_R^1),\ldots,(X^n,X_R^n))$ it holds that $(X^i\cap X_R^i)\not\subseteq Q^S_i$ for all $1\leq i\leq n$. This means there does not exist $(X^i\cap X_R^i)$ such that $(X^i\cap X_R^i)\subseteq Q^S_i$ and $(X^i\cap X_R^i)=\emptyset$. Thus, based on \Thm~\ref{them:syncTwoK} it holds that the system is $K$-step opaque.

\item $[(X^i\cap X_R^i)\subseteq Q^S_i \land (X^i\cap X_R^i)\not = \emptyset]\Rightarrow|P_{i,\Delta}(P_i(s))|>K$ for all $1\leq i\leq n$. 
Since $|P_{\Delta}(s)|\geq |P_{i,\Delta}(P_i(s))|$ and $|P_{i,\Delta}(P_i(s))|>K$ it holds that $|P_{\Delta}(s)|>K$. This means if there exists $(X^i,X_R^i)$ such that $(X^i\cap X_R^i)\subseteq Q^S_i$ and $(X^i,X_R^i)\not = \emptyset$ then $|P_{\Delta}(s)|>K$. Thus, based on \Thm~\ref{them:syncTwoK} it follows that the system $\SYSG$ is $K$-step opaque.  \QEDA
\end{itemize}

\Thm~\ref{them:OpBlockingKlor} establishes that when the transformed system is nonblocking, we can conclude that the original system is $K$-step opaque.
However, if the result of nonblocking verification is negative (blocking), then the system could still be $K$-step opaque. This problem arises for two reasons.
First, it  can be caused by the over-approximation of the state space of the two-way observer, as was explained in \Sect~\ref{sec:syncK}.
Second, it may also be caused by over-approximation of $X_{\psi_i}^i$. 
The latter may happen because $X_{\psi_i}^i$ can contain states that are not fulfilling the condition on the length of strings, $|P_\Delta(s)|>K$, in \Defn~\ref{def:kStateOpTWObs}.  
However, the length of the strings are unknown before the synchronization and synchronization may add to the length of $s$, causing fulfillment of the condition. 
%

The following example illustrates all the steps of modular $K$-step verification.

\begin{example}\label{ex:1stepOpa}
Consider the system $\SYSG=\{G_1,G_2\}$, where $G_1$ and $G_2$ are shown in \Fig~\ref{fig:notion} and \Fig~\ref{fig:syncIso}, respectively. The set of secret states of $G_1$ is $Q^S_1=\{s_1,s_3\}$ and that of $G_2$ is $Q^S_2=\emptyset$. Assume $\sync_\lor$ is used for interaction and 1-step opacity needs be verified. At the first step of the compositional approach, individual components are abstracted using opaque observation equivalence, step (i) in \Fig~\ref{fig:composAlg}. This results in the abstracted system $\tilde{\SYSG}=\{\tilde{G}_1,G_2\}$ with $\tilde{Q}^S_1=\{[s_1]\}$. $\tilde{G}_1$ is shown in \Fig~\ref{fig:notion}. Next, the two-way observers of the abstracted components are generated, step (ii) in \Fig~\ref{fig:composAlg}. As explained in \Examp~\ref{ex:twowayInfiniLor}, the two-way observer of $\tilde{G_1}$ is $\tilde{H}_1$ and the two-way observer of $G_2$ is $H_2$; shown in Figs~\ref{fig:notion} and~\ref{fig:twoSyncInf}.  Next,  1-step opacity verification is transformed to nonblocking verification by generating the $\psi_i$-automaton for each $H_i$, step (iii) in \Fig~\ref{fig:composAlg}. For the two-way observer $H_1$ we have that $X_{\psi_1}^1=\{F\}$, since for the state $F$ it holds that $\{[s_1],[s_2]\}\cap\{[s_1]\}\subseteq Q^S_1$ and $|P_{1,\Delta}((\alpha,\epsilon)(\epsilon,\beta))|=1$. For the two-way observer $H_2$ we have that $X_{\psi_2}^2=\emptyset$. \Fig~\ref{fig:nonInfini} shows  $H_{1,\psi_1}$. The system $H_{1,\psi_1}\sync H_{2,\psi_2}$ is blocking since $H_{1,\psi_1}\sync H_{2,\psi_2}\trans[(\alpha,\epsilon)(\epsilon,\beta)](F,F')\trans[\psi_1]\dumpstate$. Thus, no conclusion can be drawn about the 1-step opacity of the system, step (iv) in \Fig~\ref{fig:composAlg}. 

However, as explained in \Examp~\ref{ex:twowayInfiniLor} state $(F,F')$ is reachable due to over approximation.  Thus, it can be concluded that the system is 1-step opaque.

\end{example}

Finally, Corollary~\ref{cor:OpBlockingKland} shows how $K$-step opacity can be verified when $\sync_\land$ is used for interaction.

\def\OpBlockingKland{Let $\mathcal{G}=G_1\sync_\land\ldots\sync_\land G_n$ be a non-deterministic system, with  set of secret states $Q_{\SYSG}^S=Q_1^S\times\cdots\times Q_n^S$. Let $\Delta\colon \ACT\trans\ACT\times\{\epsilon\}$ and $\Delta_R\colon \ACT\trans\{\epsilon\}\times\ACT$ and let $H_i=\langle \ACT_H,Q_H,\trans_H,Q^\circ_H\rangle$, where $\ACT_{H_i}=\Delta(\ACT_i)\cup\Delta_R(\ACT_i)$, be the two-way observer of $G_i$  for $1\leq i\leq n$. Let $P_{i,\Delta}:\Sigma_{H_i}\to \Delta_R(\ACT_i)$.
If $H_{1,\psi}\sync \ldots \sync H_{n,\psi}$ is nonblocking then $\SYSG$ is $K$-step opaque, where where $X_{\psi}^i = \{\ H_i\trans[s](X^i, X_R^i)\ |\ (X^i\cap X_R^i)\subseteq Q_i^S \land\ (X^i\cap X_R^i)\not =\emptyset\ \land |P_{i,\Delta}(s)|\leq K\}$. }

\begin{corollary}\label{cor:OpBlockingKland}
\OpBlockingKland
\end{corollary}
\emph{Proof:}

If $H_{1,\psi}\sync \ldots \sync H_{n,\psi}$ is nonblocking it means that for all $s\in (\Delta(\ACT)\cup \Delta_R(\ACT))^*$ such that $H_{1,\psi}\sync \ldots \sync H_{n,\psi}\trans[s]((X^1,X_R^1),\ldots,(X^n,X_R^n))\not\trans[\psi]$.
This means there exists $(X^i,X_R^i)$ such that $(X^i,X_R^i)\not\in X_\psi^i$. If $(X^i,X_R^i)\not\in X_\psi^i$ then there are two following possibilities.
\begin{enumerate}
\item\label{item:KstepSyncLand1}   $(X^i\cap X_R^i)\not\subseteq Q^S_i$. This means for all $s\in (\Delta(\ACT)\cup \Delta_R(\ACT))^*$ such that $H_{1}\sync \ldots \sync H_{n}\trans[s]((X^1,X_R^1),\ldots,(X^n,X_R^n))$ it holds that there exists $(X^i,X_R^i)$ such that $(X^i\cap X_R^i)\not\subseteq Q^S_i$, which based on Corollary~\ref{col:syncTwoKland} it holds that the system is $K$-step opaque.

\item  $[(X^i\cap X_R^i)\subseteq Q^S_i \land (X^i\cap X_R^i)\not = \emptyset]\Rightarrow|P_{i,\Delta}(P_i(s))|>K$.  Since $|P_{\Delta}(s)|\geq |P_{i,\Delta}(P_i(s))|$ and $|P_{i,\Delta}(P_i(s))|>K$ it holds that $|P_{\Delta}(s)|>K$. This means if $[(X^j\cap X_R^j)\subseteq Q^S_j \land (X^j\cap X_R^j)\not = \emptyset]$ for all $1\leq j\leq n$ then $|P_{\Delta}(s)|>K$. Thus, based on Corollary~\ref{col:syncTwoKland} it follows that the system $\SYSG$ is $K$-step opaque.  \QEDA 

\end{enumerate}


\begin{example}\label{ex:kNon}
Consider the system $\SYSG=\{\tilde{G}_1,G_2\}$, where $\tilde{G}_1$ and $G_2$ are shown in \Fig~\ref{fig:notion} and \Fig~\ref{fig:syncIso}, respectively. The set of secret states of $\tilde{G}_1$ is $\tilde{Q}^S_1=\{[s_1]\}$ and that of $G_2$ is $Q^S_2=\emptyset$. Assume that $\sync_\land$ is used for interaction and that 1-step opacity needs to be verified. The two-way observer of $\tilde{G_1}$ is $\tilde{H}_1$ and the two-way observer of $G_2$ is $H_2$. Automaton $\tilde{H}_1$ is shown in \Fig~\ref{fig:notion} as $H_1$ and  $H_2$ is shown in \Fig~\ref{fig:twoSyncInf}. To transform 1-step opacity verification to nonblocking verification, the  $\psi$-automaton for each $H_i$ needs to be generated. For the two-way observer $\tilde{H}_1$ we have that $\tilde{H}_1\trans[(\epsilon,\beta),(\alpha,\epsilon)]F=\{[s_1],[s_2]\},\{[s_1]\}$ and $\{[s_1],[s_2]\}\cap\{[s_1]\}\subseteq Q^S_1$ and $|P_{1,\Delta}((\epsilon,\beta),(\alpha,\epsilon))|=1$. Thus, $X_{\psi}^1=\{F\}$ and for the two-way observer $H_2$ we have that $X_{\psi}^2=\emptyset$. Automaton $H_{1,\psi}$ is similar to $H_{1,\psi_1}$ shown in \Fig~\ref{fig:nonInfini}, 
but the transition $F\trans[\psi_1]\dumpstate$ is instead replaced by $F\trans[\psi]\dumpstate$. 
The system $H_{1,\psi}\sync H_{2,\psi}$ is trivially nonblocking. Thus, we can conclude that the system is 1-step opaque. 
\end{example}


\section{Experimental Results}\label{sec:experimental} 

\begin{figure}
	\centering
		\includegraphics{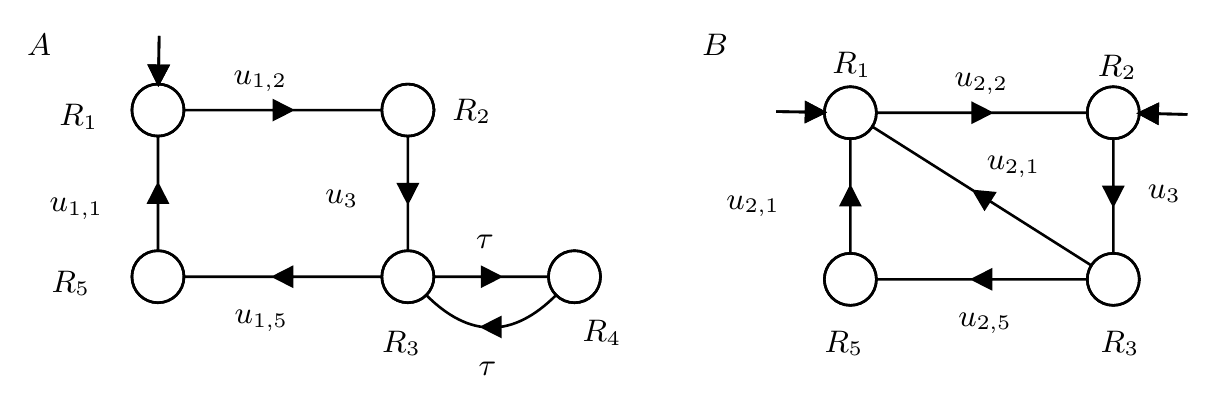}
\caption{Automata of Player $A$ and Player $B$.}\label{fig:experimental}
\end{figure}

In this section, compositional current-state opacity verification is tested on a scalable example. 
The example consists of two players, $A$ and $B$, that are moving in a house. 
To enter each room of the house the players need to use the corresponding key. 
One of the rooms is a ``two-person rule'' room, which means that to open that room the presence of the two players is required. \Fig~\ref{fig:experimental} shows the model of the players.

In the model of the players, event $u_{i,j}$ means that player $i$ has unlocked room $j$. 
Player $A$ starts at room $R_1$, from which the player can go to room $R_2$.  
Room  $R_3$ is the two-person rule room and event $u_3$ is a shared event between the two players. 
This  shows that both players can enter room $R_3$ only if they are at room $R_2$ simultaneously.
After entering room $R_3$, player $A$ can either unobservably go to room $R_4$, which is a secret room for player $A$, and then return to room $R_3$, or it can go  to room $R_5$. From room $R_5$, player $A$ can return to the initial room $R_1$. Player $B$ can start either at room $R_1$ or $R_2$. From room $R_3$,  player $B$ can go back to the initial room $R_1$ or it can go to room $R_5$ and from there go back to the initial room $R_1$. 
However, room $R_5$ is a secret room for $B$. 
The example can be scaled up by adding serially connected houses or increasing the number of  players. 


We have used Supremica~\cite{AkeFabFloVah:03} to test the scalability of our compositional methodology for current-state opacity verification   on the above example.
The model is scaled up by adding up to 2,000 players or having up to 2,000 serially connected houses (using $||_{\lor}$). 
All tests were run on a standard laptop using a quad core CPU at 2.6 GHz. 
Our methodology does not use any prior knowledge about the system. 
Table~\ref{table:experimental} shows the results of the experiments. 
For each model, the table shows the number of automata (Aut.) and whether the system is current-state opaque (Opaque). 
The table also shows the runtime for calculating opaque observation equivalence, the runtime for constructing the current-state estimator of all the components, and finally the runtime for nonblocking verification. 
Even though the compositional approach has no prior knowledge about the system, the models can be verified in few seconds or minutes.

\begin{table}[htb]
\resizebox{\columnwidth}{!}{%
\begin{tabular}{|>{\sf\bfseries}lrr|rrr|}
\hline
\multicolumn{1}{|l}{\bf Model} & \bf Aut & 
  \bf Opa. &
\multicolumn{1}{c}{\bf OOE\strut} & \multicolumn{1}{c}{\bf CSE} & 
  \multicolumn{1}{c|}{\bf Nonb.}\\
\hline
10 Players & 20 & False & 0 ms & 0 ms & 94 ms \\
100 Players & 200 & False & 0 ms & 0 ms & 359 ms \\
500 Players & $10^3$ & False & 32 ms & 16 ms & 5.5 s \\
1000 Players & 2\E 3 & False & 31 ms & 65 ms & 16.6 s \\
2000 Players & 4\E 3 & False & 63 ms & 78 ms & 64.4 s \\
\hline
10 Houses & 20 & False & 0 ms & 1 ms & 0 ms \\
100 Houses & 200 & False & 15 ms & 8 ms & 188 ms \\
500 Houses & $10^3$ & False & 62 ms & 31 ms & 1.1 s \\
1000 Houses & 2\E 3 & False & 109 ms & 62 ms & 3.9 s \\
2000 Houses & 4\E 3 & False & 152 ms & 141 ms & 11.7 s \\
\hline
\end{tabular}}%
  \caption{Experimental results}%
  \label{table:experimental}%
\end{table}

\section{Conclusion}\label{sec:conclusion} 

We introduced a novel methodology for verification of current-state and $K$-step opacity in modular discrete event systems. 
The methodology supports compositional reasoning using the notion of {opaque observation equivalence} that we defined, which  guarantees that current-state and $K$-step opacity are preserved properties. 
After abstracting the components, the opacity verification problem is then transformed to a  nonblocking verification problem. 
This makes it possible to use existing compositional methods for nonblocking verification of modular systems. 
Under our methodology, the system is current-state opaque if and only if the transformed system is nonblocking. 
In addition, it can be concluded that the system is infinite-step opaque or $K$-step opaque if the transformed system is nonblocking.  
Our experimental results suggest that the compositional approach that we have presented can lead to significant computational gains over a monolithic approach.

It would be of interest to study in the future how compositional approaches could be used to {enforce} opacity for a non-opaque system, either via supervisory control or via the use of output edit functions.

\bibliographystyle{unsrt}  


\begin{lemma}\label{lemm:blockingForLor}
Let $\SYSG=\{G_1,\ldots,G_n\}$ be a deterministic system such that the set of $\psi$-states of $G_i$ is $X^i_{\psi_i}$.  Then $\SYSG\trans[s](x_1,\ldots,x_n)$ such that there exists $x_i\in X^i_{\psi_i}$ if and only if $G_{1,\psi_1}\sync\ldots\sync G_{n,\psi_n}$ is blocking.
\end{lemma}

\emph{Proof} 
First assume $\SYSG\trans[s](x_1,\ldots,x_n)$ and there exists $x_i$ such that $x_i\in X^i_{\psi_i}$. Let $P_j:\ACT_1\cup\cdots\cup\ACT_n\trans \ACT_j$. Then based on Definition~\ref{def:synch} it holds that $G_i\trans[P_i(s)]x_i$ and $x_i\in X_{\psi_i}^i$. This means $G_{i,\psi_i}\trans[P_i(s)]x_i\trans[\psi_i]\dumpstate$. Based on Definition~\ref{def:synch} and as $G_j$ and $G_{j,\psi_j}$ for all $1\leq j\leq n$ are deterministic it holds that $G_{1,\psi_1}\sync\ldots\sync  G_{n,\psi_n}\trans[s](x_1,\ldots,x_n)\trans[\psi_i](x'_1,\ldots,x'_n)$ such that $x_i=\dumpstate$ and $x'_j=x_j$ for $j\neq i$. Now assume $(x'_1,\ldots,x'_n)\trans[t](y_1,\ldots,y_n)$. As $\dumpstate\not\trans$ in $G_{i,\psi_i}$ it holds that $P_i(t)=\varepsilon$, which implies $y_i=\dumpstate$. This means  $G_{1,\psi}\sync\ldots\sync G_{n,\psi}$ is blocking.

Now assume $G_{1,\psi_1}\sync\ldots\sync G_{n,\psi_n}$ is blocking. As all the states of $G_{j,\psi}$ for all $1,\leq j\leq n$ except $\dumpstate$ are marked, then $G_{1,\psi}\sync\ldots\sync G_{n,\psi}$ is blocking implies there exist $i$ such that $G_{1,\psi_1}\sync\ldots\sync G_{n,\psi_n}\trans[s](x_1,\ldots,x_n)\trans[\psi_i](y_1,\cdots,y_n)$ and $y_i=\dumpstate$.  Based on Definition~\ref{def:synch} it holds that $G_{i,\psi_i}\trans[P_i(s)]x_j\trans[\psi_i]\dumpstate$, which implies  $G_i\trans[P_i(s)]x_i$ and $x_i\in X_\psi^i$. Based on Definition~\ref{def:synch} and as $G_j$, $1\leq j\leq n$ are deterministic it holds $\SYSG\trans[s](x_1,\ldots,x_n)$ and $x_i\in X^i_{\psi_i}$.\QEDA




\begin{lemma}\label{lemm:blockingForAnd}
Let $\SYSG=\{G_1,\ldots,G_n\}$ be a deterministic system with the set of $\psi$-states $X^1_\psi\times \cdots\times X^n_\psi$.  Then $\SYSG\trans[s](x_1,\ldots,x_n)$ and $(x_1,\ldots,x_n)\in X^1_\psi\times \cdots\times X^n_\psi$ if and only if $G_{1,\psi}\sync\ldots\sync G_{n,\psi}$ is blocking.
\end{lemma}

\emph{Proof} 
First assume $\SYSG\trans[s](x_1,\ldots,x_n)$ and $(x_1,\ldots,x_n)\in X^1_\psi\times \cdots\times X^n_\psi$. Let $P_j:\ACT_1\cup\cdots\cup\ACT_n\trans \ACT_j$. Then based on Definition~\ref{def:synch} it holds that $G_j\trans[P_j(s)]x_j$ and $x_j\in X_\psi^j$ for $1\leq j\leq n$. This means $G_{j,\psi}\trans[P_j(s)]x_j\trans[\psi]\dumpstate$. Based on Definition~\ref{def:synch} and as $G_j$ and $G_{j,\psi}$ are deterministic it holds that $G_{1,\psi}\sync\ldots\sync  G_{n,\psi}\trans[s](x_1,\ldots,x_n)\trans[\psi]\dumpstate$, which means $G_{1,\psi}\sync\ldots\sync G_{n,\psi}$ is blocking.

Now assume $G_{1,\psi}\sync\ldots\sync G_{n,\psi}$ is blocking. As all the states of $G_{j,\psi}$, for all $1\leq j\leq n$ except $\dumpstate$ are marked, then $G_{1,\psi}\sync\ldots\sync G_{n,\psi}$ is blocking implies that $G_{1,\psi}\sync\ldots\sync G_{n,\psi}\trans[s](x_1,\ldots,x_n)\trans[\psi]\dumpstate$.  Based on Definition~\ref{def:synch} it holds that $G_{j,\psi}\trans[P_j(s)]x_j\trans[\psi]\dumpstate$, which implies  $G_j\trans[P_j(s)]x_j$ and $x_j\in X_\psi^j$ for all $1\leq j\leq n$. Based on Definition~\ref{def:synch} it holds $\SYSG\trans[s](x_1,\ldots,x_n)$ and $(x_1,\ldots,x_n)\in X^1_\psi\times \cdots\times X^n_\psi$.\QEDA

\begin{lemma}\label{lemm:renamingInsync}
Let $\SYSG$ be a deterministic system and let $\Delta :\ACT\trans \ACT\times\{\epsilon\}$. Then $\Delta(G_1\sync\ldots\sync G_n)=\Delta(G_1)\sync\ldots\sync \Delta(G_n)$.
 \end{lemma}
 
 \emph{proof} It is enough to show that $\Delta(G_1\sync\ldots\sync G_n) $ and $\Delta(G_1)\sync\ldots\sync \Delta(G_n)$ have the same transition relation. 
 
 First let $(x_1,\ldots,x_n)\trans[(\sigma,\epsilon)](y_1,\ldots,y_n)$ in $\Delta(G_1\sync\ldots\sync G_n)$. Then it holds that $(x_1,\ldots,x_n)\trans[\sigma](y_1,\ldots,y_n)$ in $G_1\sync\ldots\sync G_n$, where $\Delta(\sigma)=(\sigma,\epsilon)$. Then based on Definition~\ref{def:synch} it holds that $x_i\trans[P_i(\sigma)]y_i$ in $G_i$, which means $x_i\trans[\sigma]y_i$ if $\sigma\in \ACT_i$ and $x_i=y_i$ if $\sigma\not\in\ACT_i$. Thus, $x_i\trans[P_i((\sigma,\epsilon))]y_i$ in $\Delta(G_i)$, which implies  $(x_1,\ldots,x_n)\trans[(\sigma,\epsilon)](y_1,\ldots,y_n)$ in $\Delta(G_1)\sync\ldots\sync \Delta(G_n)$. 
 
 Now assume $(x_1,\ldots,x_n)\trans[(\sigma,\epsilon)](y_1,\ldots,y_n)$ in $\Delta(G_1)\sync\ldots\sync \Delta(G_n)$. Then it holds that $x_i\trans[P_i((\sigma,\epsilon))]y_i$ in $\Delta(G_i)$, which implies  $x_i\trans[P_i(\sigma)]y_i$ in $G_i$ for $1\leq i\leq n$. Thus, $(x_1,\ldots,x_n)\trans[\sigma](y_1,\ldots,y_n)$ in $G_1\sync\ldots\sync G_n$, which implies $(x_1,\ldots,x_n)\trans[(\sigma,\epsilon)](y_1,\ldots,y_n)$ in $\Delta(G_1\sync\ldots\sync G_n)$.\QEDA

\end{document}